\documentclass[onecolumn,aps,prd,superscriptaddress,preprintnumbers,nofootinbib,10pt]{revtex4-2}
\usepackage{amsfonts}
\usepackage[tbtags]{amsmath}
\usepackage{amssymb}
\usepackage{color}
\usepackage{courier}
\usepackage{dsfont}
\usepackage{euscript}
\usepackage{epsfig}
\usepackage{epstopdf}
\usepackage{float}
\usepackage{graphics}
\usepackage{subcaption}
\usepackage{graphicx}
\usepackage[utf8]{inputenc}
\usepackage{latexsym}
\usepackage{MnSymbol}
\usepackage{pifont}
\usepackage{physics}
\usepackage{slashed}
\usepackage{xcolor}
\usepackage[normalem]{ulem}
\usepackage{url}
\usepackage{verbatim}
\usepackage{widetable}
\usepackage{setspace}
\usepackage[plainpages=false,pdfpagelabels]{hyperref}

\allowdisplaybreaks

\hypersetup{
	colorlinks=true,
	citecolor=blue,
	citebordercolor=red,
	linktoc=all,
	linkcolor=blue,
	urlcolor=blue
}

\newcommand{\colAFV}[1]{{#1}}

\begin{document}

	\title{\bf \Large Modular Theory and the Bell-CHSH inequality \\ in  relativistic scalar Quantum Field Theory}
	
	\vspace{1cm}

	\author{J. G.A. Caribé}\email{joaogcaribe@ufrrj.br} \affiliation{ UFRRJ, Universidade Federal Rural do Rio de Janeiro, Departamento de Física, Zona Rural, BR-465, Km 07, 23890-000, Seropédica, Rio de Janeiro, Brazil}\affiliation{UERJ -- Universidade do Estado do Rio de Janeiro,	Instituto de Física -- Departamento de Física Teórica -- Rua São Francisco Xavier 524, 20550-013, Maracanã, Rio de Janeiro, Brazil}

	\author{M. S.  Guimaraes}\email{msguimaraes@uerj.br} \affiliation{UERJ -- Universidade do Estado do Rio de Janeiro,	Instituto de Física -- Departamento de Física Teórica -- Rua São Francisco Xavier 524, 20550-013, Maracanã, Rio de Janeiro, Brazil}
	
	\author{I. Roditi} \email{roditi@cbpf.br} \affiliation{CBPF $-$ Centro Brasileiro de Pesquisas Físicas, Rua Dr. Xavier Sigaud 150, 22290-180, Rio de Janeiro, Brazil}\affiliation{Institute for Theoretical Physics, ETH Z\"urich, 8093 Z\"urich, Switzerland} 
	
	\author{S. P. Sorella} \email{silvio.sorella@fis.uerj.br} \affiliation{UERJ -- Universidade do Estado do Rio de Janeiro,	Instituto de Física -- Departamento de Física Teórica -- Rua São Francisco Xavier 524, 20550-013, Maracanã, Rio de Janeiro, Brazil}

	\begin{abstract}

The Tomita-Takesaki modular theory is employed to discuss the Bell-CHSH inequality in wedge regions. By using the Bisognano-Wichmann results, the construction of a set of wedge localized vectors in the one-particle Hilbert space of a relativistic massive scalar field in $1+1$ dimensions is devised  to establish whether  violations of the Bell-CHSH inequality might occur for  different choices of Bell's operators. In particular, the construction of the wedge localized vectors employed in the seminal work by Summers-Werner is scrutinized and applied to Weyl and other operators. We also outline a possible path towards the saturation of Tsirelson's bound.

\end{abstract}
		
\maketitle

\section{Introduction}\label{sect}

The idea of the so-called modular localization is as beautiful as powerful \cite{Brunetti:2002nt,Borchers:2000pv,Schroer:2014kya}. The observation that the modular theory of Tomita-Takesaki  \cite{Takesaki:1970aki,Bratteli:1979tw,Summers:2003tf,Guido:2008jk} displays a deep connection with the Minkowski spacetime is a source of intensive investigation in many areas, ranging from Quantum Gravity to relativistic information theory, see \cite{Witten:2018zxz}. \\\\The  results of Bisognano-Wichmann in wedge regions \cite{Bisognano:1975ih}, give a transparent example of the above mentioned statements. Focusing on the case of a massive free real scalar field $\varphi(t,x)$ in $1+1$ spacetime dimensions where the quantized field can be written as
\begin{equation} 
\varphi(t,x) = \int_{-\infty}^{\infty} \frac{dk}{2\pi} \frac{1}{\omega_k} \left( e^{-i p^\mu x_\mu} a_k + e^{i p^\mu x_\mu} a^\dagger_k \right),\label{fd}
\end{equation}
where $k^\mu x_\mu = \omega_k t - k x$ with $\omega_k =\sqrt{k^2+m^2}$ and the creation and annihilation operators $a^\dagger_k$ and $a_k$ obey the canonical commutation relations~\eqref{eq:CCR}.
The Bisognano-Wichmann analysis reveals that the modular operator $\delta$ as well as the modular conjugation $j$ are related, respectively, to the self-adjoint generator of the boosts and to the $CPT$ operator\footnote{ More precisely, from \cite{Bisognano:1975ih}, it follows that 
\begin{equation} 
j= R_3(\pi) (CPT) \;, \label{br3}
\end{equation}
where $R_3(\pi)$ is a rotation of $\pi$ around the $x$-axis. The factor $R_3(\pi)$ is harmless in the case of a scalar field. However, in the Fermi case, it has to be taken into due account, see \colAFV{App.\eqref{appC}}.}, namely 
\begin{eqnarray} 
\delta & =& e^{-2\pi K}\;, \qquad K = i \frac{\partial}{\partial \theta}  \;, \nonumber \\
j &= &CPT \;\; operator \;, \label{bw}
\end{eqnarray} 
where $\theta$ stands for the rapidity variable\footnote{A boost transformation is given by 
\begin{eqnarray} 
\omega_p' & = & \omega_p \cosh(\xi) - p \sinh(\xi) \;, \nonumber \\
p' & =& p \cosh(\xi) - \omega_p \sinh(\xi) \;. \label{boosts}
\end{eqnarray} 
In rapidity space, it corresponds to a shift of $\theta$; $\theta \rightarrow \theta +\xi$.}:
\begin{equation} 
\omega_p = m \cosh(\theta) \;, \qquad p = m\sinh(\theta)  \;, \qquad \theta \in {\mathbb{R}}  \;. \label{rp}
\end{equation}
As it will be discussed in what follows, the operators $(\delta,j)$ enable us to characterize the vectors  of the 1-particle Hilbert space ${\cal H}$. These vectors  turn out to play a key role in the study of the violation of the Bell-CHSH inequality \cite{Bell:1964kc,Clauser:1969ny}. This follows from the fact that the Bell-CHSH correlator turns out to be expressed in terms of inner products of vectors belonging to the 1-particle Hilbert space \cite{Summers:1987fn,Summers:1987squ}, see Eqs.\eqref{bs} and \eqref{fg}. \\\\In relativistic Quantum Field Theory, the Bell-CHSH inequality can be formulated by following the setup outlined by  \cite{Summers:1987fn,Summers:1987squ,Summers:1987ze}, who have shown that maximal violation can be  achieved in the vacuum state $|0\rangle$ and for wedge regions, already at the level of free fields\footnote{See \cite{Guimaraes:2024byw,Guimaraes:2024mmp} for recent reviews of the Bell-CHSH inequality in Quantum Mechanics and Quantum Field Theory.}. \\\\One starts by introducing the complementary causal wedge regions $(W_R,W_L)$
\begin{equation} 
W_R =\{ (x,t)\;, \;x \ge |t| \} \;, \qquad W_L =\{ (x,t)\;, \; -x \ge |t| \} \;, \label{wrwl}
\end{equation}
and a set $\{A(f),A(f'),B(g),B(g')\}$ of Hermitian Bell operators
\begin{equation} 
A(f) = A(\varphi(f))\;, \qquad A(f') = A(\varphi(f'))\;, \qquad B(g) = B(\varphi(g))\;, \qquad B(g') = B(\varphi(g'))\;, \label{short}
\end{equation}
with  $\varphi(h)$ denoting  the smeared field  \cite{Haag:1992hx}, see Appendix \eqref{appA}.
\begin{equation} 
\varphi(h) = \int d^2x\; \varphi(x) h(x) \;, \label{smf}
\end{equation}
where $h(t,x)$ is a smooth  test function. \\\\ \colAFV{According to  eq.\eqref{wrwl}, the boundaries of the wedge regions, {\it i.e.} $x=\pm t$, have been included in the definition of $(W_R,W_L)$, although in other references, see \cite{Bisognano:1975ih}, these regions are defined as open regions, namely, without their boundaries. A few clarifying words on this issue seem to be in order. We underline that quantum fields need to be smeared out with suitable test functions $\{ h(t,x) \}$, eq.\eqref{smf}, in order to be properly defined as operators acting on the Hilbert space. In the local-algebraic construction one may take these test functions to have support compactly contained in the corresponding open wedge; equivalently, for the present discussion, they do not carry support on the boundary $x=\pm t$. Thus the inclusion or exclusion of the boundary in the symbolic notation for $(W_R,W_L)$ does not affect the smeared operators. Moreover, even though the boundary itself carries no support, regions arbitrarily close to it may contribute significantly to the Bell-CHSH correlation function. This is the case, for example, near the origin $\{x=t=0\}$, see \cite{Dudal:2023mij,Dudal:2024bmf,Dudal:2026eil}, where it is pointed out that the relevant spacetime test functions are supported in a very narrow region around $\{x=t=0\}$.} \\\\The operators $A(f),A(f'),B(g),B(g')$ are bounded,
\begin{equation} 
||A(f)|| \le 1\;, \qquad ||A(f')|| \le 1\;, \qquad ||B(g)|| \le 1\;, \qquad ||B(g')|| \le 1\;,  \label{bound}
\end{equation}
and obey the following conditions 
\begin{eqnarray} 
\left[A(f),B(g) \right] & =& 0 \;, \qquad [A(f),B(g')]=0 \;,  \qquad [A(f),A(f')] \neq 0 \;, \nonumber \\
\left[A(f'),B(g)\right] & =& 0 \;, \qquad [A(f'),B(g')]=0 \;,  \qquad [B(g),B(g')] \neq 0 \;, \label{Bella}
\end{eqnarray} 
meaning that $A(f)$ and $A(f')$ are compatible with $B(g)$ and $B(g')$. This is a consequence of the fact that the test functions $f$ and $f'$ are supported in the right wedge $W_R$, while $g$ and $g'$ are supported in the left wedge $W_L$, hence, they are supported on space-like separated regions. One then introduces the so-called Bell-CHSH correlator 
\begin{equation} 
\langle {\cal C} \rangle = \langle 0 |  (A(f) +A(f'))B(g) + (A(f)-A(f'))B(g')  |0 \rangle. \label{ineq}
\end{equation}
According to \cite{Summers:1987fn,Summers:1987squ,Summers:1987ze}, the Bell-CHSH inequality is said to be violated whenever 
\begin{equation} 
2 < \big|  \langle {\cal C} \rangle \big| \le 2 \sqrt{2} \;, \label{viol}
\end{equation} 
where the maximum value, $2\sqrt{2}$, is known as the Tsirelson's bound \cite{Cirelson:1980ry}. \\\\Concerning the Bell operators, we shall rely on a class of Hermitian bounded operators obtained as continuous superpositions of the unitary Weyl operators 
\begin{equation} 
{\cal W}_f = e^{i \phi(f)} \;, \qquad {\cal W}_f^\dagger {\cal W}_f = 1 ={\cal W}_f {\cal W}_f^\dagger \;. \label{weyl} 
\end{equation}
Following \cite{Guimaraes:2025vfu}, for the operator $A(f)$ one writes 
\begin{equation} 
A(f) = \int dk \rho(k) e^{i k \varphi(f)} \;, \label{op}
\end{equation}
where $\rho(k)$ is a suitable normalized distribution. \colAFV{Equation \eqref{op} is a shorthand notation for a rather general class of Hermitian bounded operators obtained by continuous superpositions of Weyl unitaries. This notation includes multidimensional integral representations, with distributions $\rho(k_1,\ldots,k_n)$ depending on several variables.} As an example of such operators, we may quote the bounded Hermitian operator $\Pi_f$, borrowed from  Quantum Optics, 
\begin{equation} 
A(f) = \Pi_f = \frac{1}{\pi} \int_{\mathbb{C}} d^2 \alpha \; e^{-|\alpha|^2} \; {\cal D}_f(\alpha) \qquad {\cal D}_f(\alpha) = e^{\alpha a^\dagger_f - \alpha^{*} a_f} \;, \label{par}
\end{equation} 
where ${\cal D}_f$ is the displacement operator and $(a_f,a^\dagger_f)$ are the smeared annihilation and creation operators\footnote{Given a smooth test function $h(t,x)$, the smeared annihilation and creation operators are defined as 
\begin{equation} 
a_h = \int_{-\infty}^\infty \frac{dk}{2\pi} \frac{1}{2\omega_k} h^*(\omega_k,k) a_k \;, \qquad a\dagger_h = \int_{-\infty}^\infty \frac{dk}{2\pi} \frac{1}{2\omega_k} h(\omega_k,k) a^\dagger_k \;, \qquad h(\omega_k,k) = \int d^2x e^{-i \omega_k t -kx} h(t,x) \;. \label{smaad}
\end{equation} 
}, see App.\eqref{appA}. Similar expressions hold for $A(f')$, $B(g)$ and $B(g')$. \\\\We now uncover the relationship between the Bell-CHSH correlation function $\langle {\cal C} \rangle$ and the 1-particle Hilbert space ${\cal H}$ of the theory, given by 
\begin{equation} 
{\cal H} = L^2(d\mu_p, H_m)\;, \qquad d\mu_p =\frac{dp}{2 \pi} \frac{1}{2 \omega_p} \;, \qquad H_m =\{(p_0,p)\;, p_0^2-p^2=m^2\;, p_0>0\}
\end{equation}
The inner product $<\cdot\,|\,\cdot>:L^2(d\mu_p, H_m) \to \mathbb{C}$ between two vectors $\psi_1,\psi_2 \in L^2(d\mu_p, H_m)$ is given by the Lorentz invariant expression 
\begin{equation} 
\langle \psi_1 | \psi_2\rangle = \int_{-\infty}^{\infty} \frac{dp}{2 \pi} \frac{1}{2 \omega_p} \psi_1(\omega_p,p)^* \psi_2(\omega_p,p) \;. \label{inner}
\end{equation} 
The fact that the Bell-CHSH correlation function exhibits a formulation in terms of the inner products of $L^2(d\mu_p, H_m)$ can be understood by noticing that for test functions $f,g$ with space-like separated supports one has the basic expression \cite{Guimaraes:2024mmp}. 
\begin{equation} 
\langle 0|\; e^{i \varphi(f)} e^{i \varphi(g)}\;|0\rangle = e^{-\frac{1}{2} \langle \psi_f +\psi_g| \psi_f +\psi_g\rangle } \;, \label{bs}
\end{equation}
where $\psi_f$ and $\psi_g$ are given by 
\begin{equation} 
\psi_f(\omega_p,p) = \int d^2x \; e^{-i (t \omega_p  - x p)} f(t,x) \;, \qquad 
\psi_g(\omega_p,p) = \int d^2x \;e^{-i (t \omega_p  - x p)} g(t,x) \;,\label{fg}
\end{equation}
with $supp(f) \subset W_R$ and $supp(g) \subset W_L$. As emphasized in \cite{Guido:2008jk}, expressions \eqref{fg} define an embedding of the space of the test functions into the 1-particle Hilbert space $L^2(d\mu_p, H_m)$. Equations \eqref{fg} can be seen as the restriction  of the usual Fourier transform to the mass hyperboloid $H_m$. \\\\From expression \eqref{fg}, it becomes apparent that our ability in detecting possible violations of the Bell-CHSH inequality relies ultimately in our expertise in finding a suitable set of vectors $\{\psi_f, \psi_{f'}\}$  and $\{\psi_g, \psi_{g'}\}$ and in controlling their inner products. \\\\The aim of the present work is twofold. First, we  restate the pivotal role that the modular theory plays in the analysis of the Bell-CHSH inequality in Quantum Field Theory. Second, we outline  how the modular operators $\delta$ and $j$ can be employed, in practice,  to build a suitable set of vectors $\{ \psi \in  L^2(d\mu_p, H_m)\}$ giving rise to a helpful framework in order to search for  violations of the Bell-CHSH inequality. \\\\It is worth mentioning here that, unlike the bosonic case, the issue of Bell-CHSH inequality for Fermi fields exhibits remarkable differences. In the fermionic case, the anti-commutation relations enable one to introduce in a rather simple way dichotomic operators \colAFV{built} out directly from the smeared spinor field \cite{Summers:1987fn,Summers:1987squ,Summers:1987ze}. When combined with the modular theory, this feature provides a clean proof of the saturation of the Tsirelson bound in the vacuum state, within a Quantum Field Theory framework. To our knowledge, a similar treatment in the Bose case is not yet at our disposal. As such, besides the construction of a suitable set of vectors in the 1-particle Hilbert space, we also call attention to a set of field operators, built from the scalar field,  which might give rise to maximal violation of the Bell-CHSH inequality~\eqref{viol}. Specifically, the so-called vertex operators arising in the bosonization of $(1+1)$ models \cite{Coleman:1985rnk} could offer a concrete and viable example of operators to be exploited in order to saturate Tsirelson's bound. Needless to say, the vertex operators behave like fermions, allowing thus for a helpful bridge with the Fermi case. \\\\The work is organized as follows. In Sect.\eqref{sect2} we provide the details of the construction of the vectors $\{ \psi\}$ by means of $(\delta,j)$. Sect.\eqref{sect3} is devoted to the applications to the violation of the Bell-CHSH inequality. In Sect.\eqref{sect4} we revise the construction of a set of vectors employed by Summers-Werner  \cite{Summers:1987fn,Summers:1987squ,Summers:1987ze} in their proof on the maximal violation of the Bell-CHSH inequality. Further, we pinpoint the necessity of relying on Hermitian bounded operators which encode information about the spectrum of the modular operator $\delta$. As we shall see, this requirement turns out to be a key feature to reach violations close to Tsirelson's bound $2\sqrt{2}$. Section \eqref{sect5} contains our conclusion. In  Appendix \eqref{appA} a brief account on the canonical quantization of the scalar field is provided; \colAFV{Appendix \eqref{appB} gives a short account of the one-particle Reeh-Schlieder property used in Sect.\eqref{sect4}; and Appendix \eqref{appC} revises the saturation of the Tsirelson bound for free Fermi fields, taking as an example a massless Majorana spinor.}

\section{Construction of the vectors $\{ \psi \} \in L^2(d\mu_p, H_m) $ by employing the modular operators $(\delta,j)$}\label{sect2}

\subsection{Background}

In order to present the construction of the vectors $\{ \psi \} \in L^2(d\mu_p, H_m) $ by means of  the modular operators $(\delta,j)$, a brief account of some useful mathematical tools is in order. We shall follow the extensive review by \cite{Guido:2008jk}. \\\\The first tool to be noticed is the use of the  rapidity variable $\theta$, Eq.\eqref{rp}. From,
\begin{equation} 
\frac{dp}{\omega_p} = d \theta \;, \label{ipt}
\end{equation} 
it follows that the inner product \eqref{inner} becomes 
\begin{equation} 
\langle \psi_1 | \psi_2\rangle = \int_{-\infty}^{\infty} d\theta \; \psi_1(\theta)^* \psi_2(\theta) \;, \label{ith}
\end{equation}
where we have removed the irrelevant factor $1/(4\pi)$. The 1-particle Hilbert space is thus $L^2(\theta, \mathbb{R})$. \\\\Let us consider  now the vector $\psi_f(\theta) \in L^2(\theta, \mathbb{R})$ 
\begin{equation} 
\psi_f(\theta) = \int d^2x \; e^{-im ( t \cosh(\theta)- x \sinh(\theta))} f(t,x) \;,  \label{str1}
\end{equation}
where $f(t,x)$ is real and supported in $W_R$, $supp(f) \subset W_R$. It turns out that $\psi_f(\theta)$ admits a bounded analytic extension in the strip $\{ z= \theta + i y, \; 0 \le y \le \pi\}$. In fact, we have 
\begin{equation} 
\psi_f(\theta+i y) = \int d^2x\; e^{-im t \cosh(\theta+iy) - x \sinh(\theta+iy)} f(t,x) \;. \label{str2}
\end{equation}
Looking then at the real part of the exponential, one finds 
\begin{equation} 
e^{-m \sin(y) (x \cosh(\theta) - t \sinh(\theta))} \;. \label{str3}
\end{equation}
Since, in the right wedge $W_R$, $x\ge|t|$, one sees that the real part is bounded for $0 \le y \le \pi$, as $\sin(y) \ge0$ in this interval. This property will be relevant in order to define the domain of the operator $\delta$. \\\\Following \cite{Guido:2008jk}, one introduces the real\footnote{ We remind that in a real vector space only combinations with real coefficients are allowed. Namely, if $\psi_1$ and $\psi_2$ are elements of a real vector space, only linear combinations $\alpha \psi_1+ \beta \psi_2$ with $\alpha, \beta \in \mathbb{R}$, are allowed. Despite that fact, $\psi_1$ and $\psi_2$ may be complex valued.} linear closed vector space $K(W_R)$: 
\begin{equation} 
K(W_R) =\{ \psi_f(\theta)\;, \; \psi_f(\theta)= \int d^2x\; e^{ -i (t \omega_p(\theta) - x p(\theta))} f(t,x)\;, \; supp(f) \subset W_R \} \;. \label{KWR}
\end{equation}
It is also useful to introduce the space  $K'(W_R)$ defined as 
\begin{equation} 
K'(W_R) = \{ h \in L^2(\theta, \mathbb{R})\;, \;\Im \langle h|k \rangle=0 \;, \;\forall k \in K(W_R) \} \;. \label{KpWR}
\end{equation}
It is not difficult to see that $K'(W_R) $ encodes the principle of causality, as expressed by the Pauli-Jordan distribution $\Delta_{PJ}(x-x')$, App.\eqref{appA}. For that, it suffices to rewrite the inner product  \eqref{inner} in configuration space:
\begin{equation}
\langle \psi_f | \psi_g\rangle = \int d^2x d^2x'  f(x) H(x-x') g(x') + i \int d^2x d^2x'f(x) \Delta_{PJ}(x-x') g(x') \;, \label{iconf}
\end{equation}
where the kernels $H(x-x')$ and $\Delta_{PJ}(x-x')$ are known, respectively, as the Hadamard and Pauli-Jordan distributions: 
\begin{eqnarray} 
H(x-x') & = & \int_{-\infty}^{\infty} \frac{dp}{2 \pi} \frac{1}{2 \omega_p} \; \cos\left( \omega_p(t-t') - p (x-x') \right) \;, \nonumber \\
\Delta_{PJ}(x-x') & = & -\int_{-\infty}^{\infty} \frac{dp}{2 \pi} \frac{1}{2 \omega_p} \; \sin\left( \omega_p(t-t') - p (x-x') \right) \;, \label{HDPJ}
\end{eqnarray}
Both $H(x-x')$ and $\Delta_{PJ}(x-x')$ are Lorentz invariant. Moreover, $\Delta_{PJ}(x-x')$ vanishes for spacelike separations, a feature which expresses the principle of relativistic causality. \\\\We should recall here that $W_L$ is precisely the causal complement of $W_R$ and vice-versa. Given an open bounded region ${\cal O}$ of the Minkowski spacetime, its causal complement is given by ${\cal O}'$ 
\begin{equation} 
{\cal O}'= \{ x^\mu \in {\mathbb{R}}^2\;, (x-y)^2 < 0\;, \forall y^\mu \in {\cal O} \} 
\end{equation}
Thus 
\begin{equation} 
W'_R = W_L \;. \label{WRWL}
\end{equation}
The relevance of the vector spaces $K(W_R)$ and $K'(W_R)$ is due to a result by Araki \cite{Araki}, see  \cite{Guido:2008jk}, stating that 
\begin{itemize} 
\item 
\begin{equation} 
K(W'_R) = K(W_L) = K'(W_R) \;, \label{dual}
\end{equation}

\item 
\begin{equation} 
K(W_R) \cap i K(W_R) = \{ 0 \} \;, \label{empty} 
\end{equation}

\item 
\begin{equation} 
K(W_R) + i K(W_R) \;\;\; \mathrm{ is \; dense  \, \label{dense} \; in \; the \; Hilbert \; space}\,L^2(\theta, \mathbb{R}).
\end{equation}

\end{itemize}
Property \eqref{dual} is known as Haag's duality \cite{Haag:1992hx}. Property \eqref{empty}  expresses the fact that no nontrivial vectors exist belonging simultaneously to $K(W_R)$ and $i K(W_R)$. Finally, equation \eqref{dense} expresses the important fact that any vector $\psi_f(\theta) \in L^2(\theta, \mathbb{R})$ can be arbitrary well approximated by sequences of elements of $K(W_R)$ and $i K(W_R)$. \\\\A real vector space enjoying properties \eqref{empty}, \eqref{dense} is known as a standard subspace \cite{Guido:2008jk}. \\\\As shown by \cite{Rieffel}, for a standard subspace like $K(W_R)$ it is possible to set a modular theory. One introduces an unbounded\footnote{Being unbounded, care needs to be taken to properly define the domain of $s_k$. As it will be discussed later on, the domain of $s_k$ will be identified with the set of vectors $\{ \psi \} \in L^2(\theta, \mathbb{R}) $ which admit a bounded analytic continuation in the strip $z=\theta +iy, 0<y<\pi$, see Eqs.\eqref{scd} and \eqref{scd1}.  } anti-linear operator $s_{K}$, $s^2_K=1$, whose action on $K(W_R)+i K(W_R)$ is specified by 
\begin{equation} 
s_K (h+i k) = h -i k \;, \qquad h,k \in K(W_R) \;, \label{acts}
\end{equation}
The modular operators $(\delta_K,j_K)$ are thus introduced by means of the polar decomposition of $s_K$, namely 
\begin{equation} 
s_{K} = j_{K} \delta^{1/2}_K \;. \label{jde}
\end{equation}
The modular operator $\delta_K$ is self-adjoint, while $j_K$ is anti-unitary. The following properties hold \cite{Guido:2008jk}: 
\begin{eqnarray} 
j^2_K &=& 1 \;, \qquad j_K \delta_K^{1/2} j_K = \delta_K^{-1/2} \;, \nonumber \\
s_K^\dagger & = & j_K \delta^{-1/2}_K \;, \qquad \delta_K = s^\dagger_K s_K \;. \label{ssd}
\end{eqnarray}
In such a situation the Tomita-Takesaki theorem states that \cite{Guido:2008jk} 
\begin{itemize} 
\item 
\begin{equation} 
s_{K'} = s^\dagger_K \;, \label{tt1}
\end{equation} 

\item
\begin{equation} 
j_K K(W_R) = K'(W_R)  \;, \label{tt2} 
\end{equation}

\item 
\begin{equation} 
\delta^{i t}_K K(W_R) = K(W_R) \;, \qquad t\in \mathbb{R} \;, \label{tt3}
\end{equation}

\item
\begin{equation} 
 K(W_R) \cap  K'(W_R) = \{ h\;, j_K h=h \;, \delta_K h=h \}  \;. \label{tt4} 
\end{equation}
\end{itemize} 
In particular, property \eqref{tt2} expresses the important fact that the modular conjugation $j_K$ exchanges $K(W_R)$ by its complement $K'(W_R)$. From Eq.\eqref{tt1} one learns that the operator $s_{K'}$ corresponding to $K'(W_R)$ is nothing but $s_K^{\dagger}$. Also, Eq.\eqref{tt3} expresses the invariance of $K(W_R)$ under the unitary modular flow $\delta^{it}_K = e^{i t \log(\delta_K)}$. Finally, property \eqref{tt4} expresses the fact that  $K(W_R) \cap  K'(W_R) $ might be non-vanishing. Vectors belonging to $ K(W_R) \cap  K'(W_R) $ are called edge vectors. \\\\From now on, we remove the index $K$, employing the simpler notation $(j,\delta)$. Reference to $(W_R,W_L)$ will be always implicit. \\\\It turns out  \cite{Guido:2008jk} that the Tomita-Takesaki theorem enables one to characterize the standard subspace $K(W_R)$ as 
\begin{equation} 
K(W_R) = \{ h \in L^2(\theta, \mathbb{R})\;, s h =h \} \;. \label{chst}
\end{equation}
Analogously 
\begin{equation} 
K'(W_R) = \{ k \in L^2(\theta, \mathbb{R})\;, s^\dagger k =k \} \;. \label{chstd}
\end{equation}
At this stage, one takes Eqs.\eqref{chst},\eqref{chstd} as the basic tools for the modular localization. According to \cite{Guido:2008jk}, one says that a vector $\psi(\theta)$ is wedge localized if 
\begin{eqnarray} 
s \psi(\theta) & = & \psi(\theta) \;, \qquad localization \; in \; W_R \;, \nonumber \\
s^\dagger \psi(\theta) & = & \psi(\theta) \;, \qquad localization \; in \; W_L \;. \label{locz}
\end{eqnarray}
These two equations summarize the beautiful idea of the modular localization. As we shall see in the next section, they can be employed to construct  a suitable set of vectors $\{ \psi \}$.  \\\\Let us conclude this brief account by showing that $\Im \langle h|k\rangle =0$ for $h\in K(W_R)$ and $k \in K(W_L)$. In fact, from 
\begin{equation} 
s h = h \;, \qquad s^{\dagger} k = k \;, \label{ssdhk}
\end{equation}
we have 
\begin{equation} 
\langle h|k\rangle = \langle sh | s^\dagger k \rangle = \langle k|s^2 h\rangle = \langle k | h\rangle = \langle h|k\rangle^*\;, \label{ppff}
\end{equation} 
where the anti-linearity of $s$ and $s^\dagger$ has been taken into account by using $\langle s h | \psi \rangle = \langle s^{\dagger} \psi | h \rangle$. 

\subsection{The Bisognano-Wichmann results and the construction of  modular localized vectors}\label{mdl}

As already mentioned, the Bisognano-Wichmann results \cite{Bisognano:1975ih} state that, for wedge regions
\begin{equation} 
\delta = e^{-2\pi K}\;, \qquad K =i \frac{\partial }{\partial \theta} \;, \qquad j= CPT \;. \label{bbww}
\end{equation} 
From these expressions one might deduce the action of the anti-linear operator $s$ on a generic vector $\psi(\theta)$., namely 
\begin{eqnarray} 
\delta^{1/2} \psi(\theta) &= & \psi(\theta - i \pi) \;, \nonumber \\
j \psi(\theta) & = & \psi(\theta)^* \;. \label{adj}
\end{eqnarray}  
Thus 
\begin{equation} 
s \psi(\theta) = (\psi(\theta - i \pi))^* \;. \label{scd}
\end{equation}
The condition of modular localization $s \psi =\psi$ becomes then 
\begin{equation} 
\psi(\theta) = (\psi(\theta - i \pi))^* \;.  \label{scd1}
\end{equation}
Vectors fulfilling this condition are constructed by observing that the operator $(1+s)/2) $ is a projector 
\begin{equation} 
s(1+s) = (1+s)  \;. \label{pjpj}
\end{equation}
As such, a wedge $W_R$-localized vector $\omega(\theta)$ is obtained by setting 
\begin{equation} 
\omega(\theta) = \frac{(1+s)}{2} \phi(\theta) \;, \label{omt}
\end{equation} 
for some suitable $\phi(\theta)$. Evidently 
\begin{equation} 
s \omega(\theta) = \omega(\theta) \;. \label{ominv}
\end{equation} 
Though, a few words of caution about the choice of $\phi(\theta)$ are in order. As the operator $s$ is unbounded, care should be taken about its domain \ \cite{Guido:2008jk}. Here, it is worth reminding properties  \eqref{str1}-\eqref{str3}. Accordingly, one demands that $\phi(\theta)$ exhibits an analytic bounded extension in the strip $\{ z=\theta +iy\;, 0 \le y \le \pi \}$. In that way, equation \eqref{scd1} is well defined as well as the action of $(s,\delta,j)$. \\\\We are now ready to give some examples of vectors fulfilling the wedge localization condition \eqref{scd1}. Observing that 
\begin{equation} 
\cosh(\theta - i \pi) = - \cosh(\theta) \;, \qquad \sinh(\theta - i \pi) = - \sinh(\theta)\;, \label{chsh}
\end{equation}
we might take 
\begin{equation} 
\phi(\theta) = e^{-\theta^2} P(\cosh^2(\theta)) \;, \label{ex}
\end{equation}
where $P(\cosh^2(\theta))$ is an arbitrary polynomial in $\cosh^2(\theta)$, namely 
\begin{equation} 
P(\cosh^2(\theta) = \sum_{i=0}^n c_i\; (\cosh^2(\theta))^{2i} \;, \label{pol}
\end{equation}
for some free coefficients $c_i$. As one can easily figure out, the exponential factor $e^{-\theta^2}$ ensures the existence of the required analytic extension in the strip $\{z=\theta + iy \;, 0\le y \le \pi \}$. For simplicity, in Eq.\eqref{pol}, we have taken only even powers of $\cosh(\theta)$, due to $\cosh(\theta-i\pi)^2 = \cosh(\theta)^2$. However, odd powers may be envisaged too. \\\\For the vector $\omega(\theta)$ we get 
\begin{equation} 
\omega(\theta) = \frac{(1+ j \delta^{1/2})}{2} \phi(\theta) = \frac{1}{2} P(\cosh^2(\theta)) e^{-\theta^2} \left( 1+ e^{\pi^2} e^{-2\pi i \theta} \right)  \;. \label{ex1} 
\end{equation}
According now to Tomita-Takesaki theorem, a vector ${\tilde \omega}(\theta)$ localized in $W_L$ can be obtained from $\omega(\theta)$ by applying the modular conjugation $j$, {\it i.e} 
\begin{equation} 
{\tilde \omega}(\theta) = j \;\omega(\theta) \;. \label{jjom}
\end{equation}
In fact 
\begin{equation} 
s^{\dagger} {\tilde \omega}(\theta) = j \delta^{-1/2} j \;\omega(\theta) = \delta^{1/2} \omega(\theta)=  j s \omega(\theta) = j\omega(\theta) = {\tilde \omega}(\theta) \;. \label{sddjom}
\end{equation} 
Explicitly 
\begin{equation}
{\tilde \omega}(\theta)= \frac{1}{2} P(\cosh^2(\theta)) e^{-\theta^2} \left( 1+ e^{\pi^2} e^{2\pi i \theta} \right) \;. \label{exptom}
\end{equation}
It is now a simple calculation to check that 
\begin{equation} 
\Im \langle \omega | {\tilde \omega} \rangle = 0 \;. \label{zc}
\end{equation}
We underline that many choices for the polynomial $P(\cosh^2(\theta))$ are available: Hermite polynomials, hyperbolic harmonics, wavelets, etc. \\\\As we shall see, vectors like those of Eq.\eqref{ex1} can be successfully employed to investigate the violation of the Bell-CHSH inequality. We can make use of as many free coefficients $c_i$ as we wish and fine tune them to detect sensible violations. This will be the task of the next section.

\section{Applications to the Bell-CHSH inequality}\label{sect3}

\begin{figure}[t!]
\begin{minipage}[b]{0.4\linewidth}
\includegraphics[width=\textwidth]{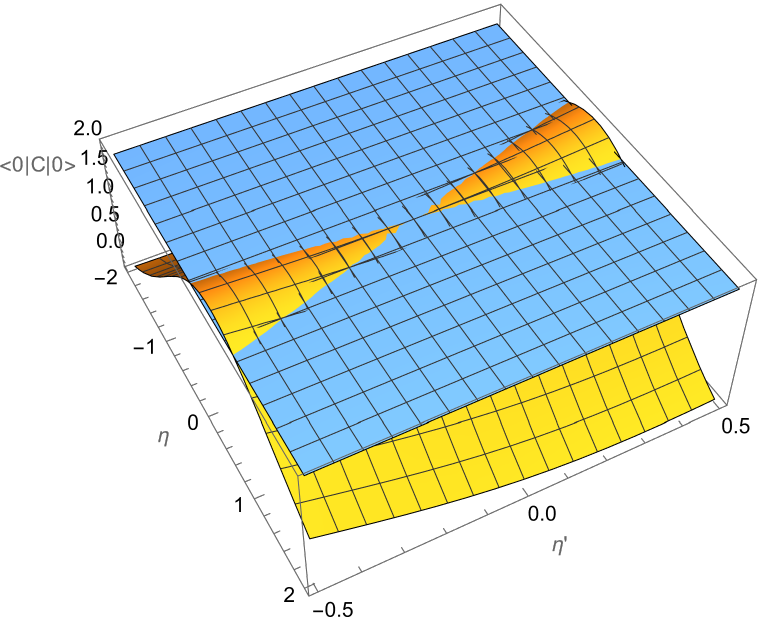}
\end{minipage} \hfill
 \caption{Behavior of the Bell-CHSH correlator $\langle 0 |{\cal C} |0\rangle$ from equation~\eqref{Bccc} as function of the free parameters $\eta$ and $\eta'$. The orange surface above the blue one corresponds to the regions where the violation takes place.}
\label{Weyl1}
\end{figure}

\begin{figure}[t!]
\begin{minipage}[b]{0.4\linewidth}
\includegraphics[width=\textwidth]{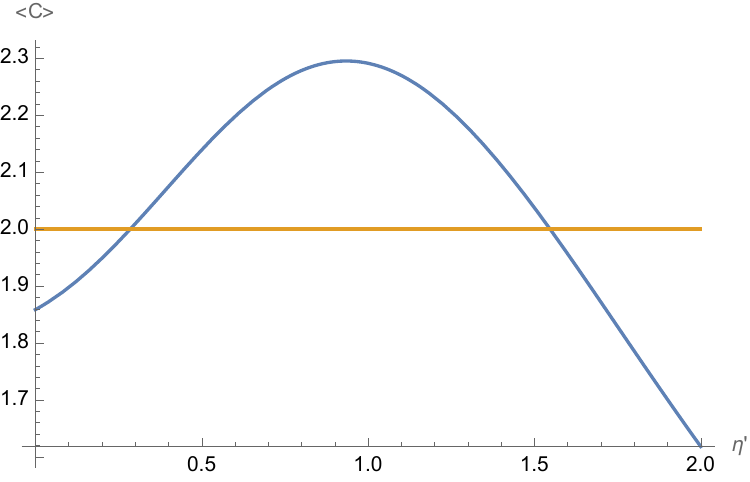}
\end{minipage} \hfill
 \caption{Behavior of the Bell-CHSH correlator  $\langle 0|{\cal C} |0\rangle$ as function of the parameter $\eta'$, for $\eta= -0.395$. One sees that the size of the violation is of about $\approx 2.3$}
\label{Weyl2}
\end{figure}

Let $\psi(\theta)$ a $W_R$ localized vector, $s\psi=\psi$. For the corresponding unitary Weyl operator ${\cal A}_{\psi}$ , Eq.\eqref{weyl},  we may write
\begin{equation} 
{\cal A}_\psi = e^{i (a_\psi + a^\dagger_\psi) } \;,  \label{weylpsi} 
\end{equation}
where 
\begin{eqnarray} 
a_\psi & = & \int d \theta \; \psi(\theta)^* a_\theta \;, \nonumber \\
a^\dagger_\psi & = & \int d \theta \; \psi(\theta) a^\dagger_\theta \;, \label{aadth}
\end{eqnarray}
are the smeared creation and annihilation operators in rapidity space 
\begin{equation} 
\left[ a_\theta, a^\dagger_{\theta'} \right]= \delta(\theta-\theta') \;,  \qquad \left[ a_\theta, a_{\theta'} \right]= \left[ a^\dagger_\theta, a^\dagger_{\theta'} \right]=  0 \;, \label{ccr}
\end{equation}
It follows that 
\begin{equation} 
\left[ a_{\psi_1}, a^\dagger_{\psi_2} \right] = \langle \psi_1 | \psi_2 \rangle \;. \label{inaad}
\end{equation}
Making use of the Baker-Campbell-Hausdorff formula, one checks that 
\begin{equation} 
{\cal A}_h {\cal A}_k = e^{-\frac{i}{2}\Im\langle h|k\rangle} {\cal A}_{(h+k)} \label{ww1} 
\end{equation}
as well as
\begin{equation} 
\langle 0| {\cal A}_h|0\rangle = e^{-\frac{1}{2} ||h||^2}  \;. \label{ww2} 
\end{equation}
Thus 
\begin{equation} 
\langle 0| {\cal A}_h {\cal A}_k |0 \rangle = e^{\frac{-1}{2} ||h+k||^2} \;, \qquad h \in K(W_R)\;, k \in K(W_L) \;. \label{ww3}
\end{equation}
As observed in \cite{Guimaraes:2024alk}, the reality properties of the correlation functions of the Weyl operators, as expressed by Eq.\eqref{ww3}, \colAFV{enable us to employ} the Weyl operators directly for a test of the Bell-CHSH inequality. Setting 
\begin{equation} 
{\cal C} = \left( {\cal A}_\psi + {\cal A}_{\psi'} \right) {\cal A}_{j\psi} +  \left( {\cal A}_\psi - {\cal A}_{\psi'} \right) {\cal A}_{j\psi'} \;, \label{Bapsi}
\end{equation}
for the Bell-CHSH correlation function $\langle 0| {\cal C}|0\rangle $ one gets 
\begin{equation} 
\langle 0| {\cal C}|0\rangle = e^{-\frac{1}{2}||\psi + j\psi||^2 } + \colAFV{e^{-\frac{1}{2}||\psi' + j\psi||^2 }} + e^{-\frac{1}{2}||\psi + j\psi' ||^2 } -e^{-\frac{1}{2}||\psi' + j\psi' ||^2 }  \;. \label{Bccc}
\end{equation}
The choice of $\psi, \psi' \in K(W_R)$ proceeds now as outlined previously. We shall use vectors with a minimal number of free parameters: $(\eta, \eta')$ 
\begin{eqnarray} 
\psi(\theta) & = & \frac{\eta}{2} (1+ j \delta^{1/2}) \;e^{-\theta^2} \;, \nonumber \\
\psi'(\theta)& = & \frac{\eta'}{2} (1+ j \delta^{1/2}) \;e^{-\theta^2}\cosh^2(\theta)  \;.  \label{minm}
\end{eqnarray}
Even with such a reduced number of parameters, it is already possible to detect a rather nice violation, as reported in Figs.\eqref{Weyl1},\eqref{Weyl2}. The maximum value of the violation is found to be 
\begin{equation} 
\langle 0 | {\cal C} | 0\rangle = 2.295 \;. \label{viol23}
\end{equation}

\section{Construction of the Summers-Werner vectors in rapidity space}\label{sect4}

This section is devoted to the study of the four vectors $(f,f',g,g')$ employed by Summers-Werner in their proof on the existence of the maximal violation of the Bell-CHSH inequality in wedge regions  \cite{Summers:1987fn,Summers:1987squ,Summers:1987ze}. \\\\Let us start by reminding that, for wedge regions, the generator of the boosts is $K = i \frac{\partial}{\partial \theta}$. \colAFV{The eigenstates} of $K$ are plane waves in rapidity space, {\it i.e.} 
\begin{equation} 
K \psi_\omega(\theta)  =  \omega \;\psi(\theta) \;,\qquad 
\psi_{\omega}(\theta) =  \frac{e^{-i \omega \theta}}{\sqrt{2 \pi}} \;, \qquad \langle\psi_\omega|\psi_{\omega'}\rangle = \delta(\omega -\omega') \;, \qquad (\omega,\omega') \in \mathbb{R} \;,   \label{eig}
\end{equation}
As a consequence, the spectrum of $\delta$ is 
\begin{equation} 
\delta = e^{-2 \pi K} \;, \qquad \delta \psi_\omega = e^{-2 \pi \omega} \psi_\omega \;, \label{eigd}
\end{equation}
with eigenvalues $\lambda^2 = e^{- 2\pi \omega}$, with $\lambda \in [0,\infty]$. \\\\To construct the four Summers-Werner vectors $(f,f')$, $(g,g')$ one introduces a normalized eigenstate $\Phi$ of the modular operator $\delta$ belonging to the spectral interval $[\lambda^2 - \epsilon,\lambda^2+ \epsilon]$, $\lambda^2 \in [0,1]$ \footnote{\label{footnote6}The appearance of the spectral interval $\lambda^2 \in [0,1]$ has quite deep reasons, see  \cite{Summers:1987ze}. It stems from the observation that the operator algebra of a Quantum Field Theory is a von Neumann algebra of the type $III_\lambda$, with $\lambda \in [0,1]$. Moreover, the endpoint $\lambda=1$ has a very special meaning, being the fixed point of the modular flux $\delta^{i t} = e^{i t \log(\delta)}$. As shown in \cite{Summers:1987fn,Summers:1987squ,Summers:1987ze}, maximal violation of Bell-CHSH inequality occurs for $\lambda \approx 1$}. Such an eigenstate can be obtained by means of a narrowed wave packet localized around $\lambda^2$, namely 
\begin{equation} 
\Phi(\theta) = {\cal N} \int_{-\infty}^\infty d\omega \; e^{-\frac{(\omega-\lambda^2)^2}{2 \epsilon}} e^{-i \omega \theta}  \;, \label{eigP}
\end{equation}
with ${\cal N}$ the normalization factor ensuring that $\langle \Phi|\Phi\rangle =1$:
\begin{equation} 
{\cal N} = \left( 2\pi \int_{-\infty}^\infty d\omega \; e^{-\frac{(\omega-\lambda^2)^2}{ \epsilon}}\right)^{-1/2} \;. \label{NN}
\end{equation}
As one can easily figure out, the exponential factor $e^{-\frac{(\omega-\lambda^2)^2}{2 \epsilon}}$ acts as a kind of spectral projector on the desired interval $[\lambda^2-\epsilon, \lambda^2+ \epsilon]$. The four Summers-Werner vectors are thus given by 
\begin{eqnarray} 
f & =& (1 -\mu_{+})^{-1/2} (1+s) \Phi\;, \qquad f'  = (1 -\mu_{+})^{-1/2} (1+s) i\Phi\; \nonumber\;, \qquad sf=f \;, \qquad sf'= f' \;, \\
g & =& (\mu_{-}-1)^{-1/2} (1+s^\dagger) \Phi\;, \qquad g'  = -(\mu_{-}-1)^{-1/2} (1+s^\dagger) i\Phi\;, \qquad s^\dagger g = g \;, \qquad s^\dagger g'= g' \;, \label{fsw}
\end{eqnarray} 
with 
\begin{equation} 
\mu_{+} = \langle \Phi | \delta | \Phi\rangle \underset{\epsilon \rightarrow 0}\rightarrow \lambda^2 \;, \qquad \mu_{-} = \langle \Phi | \delta^{-1} | \Phi\rangle \underset{\epsilon \rightarrow 0}\rightarrow \lambda^{-2} \;. \label{epsl}
\end{equation}
The vectors $(f,f')$ are localized in $W_R$, while $(g,g')$ in $W_L$. These vectors turn out to fulfill the following conditions 
\begin{eqnarray} 
||f||^2 & = & ||f'||^2 = ||g||^2 =||g'||^2 = \colAFV{\frac{1+\lambda^2}{1-\lambda^2}} \;, \qquad 
\langle f | f'\rangle  =  \langle g | g'\rangle = i \;, \nonumber \\
\langle f | g \rangle & = & - \langle f'| g'\rangle = \frac{2 \lambda}{1- \lambda^2} \;, \qquad 
\langle f | g'\rangle  =  \langle f'| g \rangle = 0 \;, \label{SWinnp}
\end{eqnarray}
\colAFV{Let us write $(f,f')$ and $(g,g')$ in explicit form, namely} 
\begin{eqnarray} 
f & = & (1 -\mu_{+})^{-1/2} {\cal N} \int d\omega \; e^{-\frac{(\omega-\lambda^2)^2}{2 \epsilon}} \left( e^{-i \omega \theta} + e^{i \omega \theta} e^{-\pi \omega} \right) \;, \nonumber \\
f' & = &i  (1 -\mu_{+})^{-1/2} {\cal N} \int d\omega \; e^{-\frac{(\omega-\lambda^2)^2}{2 \epsilon}} \left( e^{-i \omega \theta} - e^{i \omega \theta} e^{-\pi \omega} \right) \;, \nonumber \\
g & = & (\mu_{-}-1)^{-1/2}{\cal N} \int d\omega \; e^{-\frac{(\omega-\lambda^2)^2}{2 \epsilon}} \left( e^{-i \omega \theta} + e^{i \omega \theta} e^{\pi \omega} \right) \;, \nonumber \\
g' & = - i & (\mu_{-}-1)^{-1/2}{\cal N} \int d\omega \; e^{-\frac{(\omega-\lambda^2)^2}{2 \epsilon}} \left( e^{-i \omega \theta} - e^{i \omega \theta} e^{\pi \omega} \right) \;. 
\end{eqnarray}
All integrals are of a Gaussian type, being easily worked out. \colAFV{As outlined in \cite{Summers:1987fn,Summers:1987squ}, expressions \eqref{SWinnp} are very helpful in the proof of the existence of maximal violation of the Bell-CHSH inequality for both bosonic and Fermi free fields in wedge regions. This stems from the fact that Eqs.\eqref{SWinnp} depend directly on the modular structure of the theory, encapsulated in the spectral parameter $\lambda$. Let us conclude this section by remarking that, as previously mentioned, the real subspace $K(W_R) $ is a standard subspace, meaning, in particular, that $K(W_R) + iK(W_R)$ is dense in the 1-particle Hilbert space. This important property, valid in fact for any open non-empty region $O$ of the Minkowski spacetime, is inherited from the Reeh-Schlieder theorem, being referred to as the Reeh-Schlieder theorem for the 1-particle Hilbert space, see \cite{Guido:2008jk}. As a consequence of this result, a wave packet such as that of eq.\eqref{eigP} need not be the exact image of a single compactly supported test function, but can be arbitrarily well approximated by a suitable sequence of spacetime test functions supported in $W_R$, as expressed by eq.\eqref{KWR}. Due to the relevance of this result, a sketch of the proof is reported in App.\eqref{appB}.} 

\subsection{Applications} 

\subsubsection{Weyl operators revisited} 
As a first application of the previous construction, we revise here the correlation function of pure Weyl operators, Eq.\eqref{weylpsi}. This example will enable us to highlight a few aspects related to the four vectors $(f,f')$ and $(g,g')$. in fact, as one sees from Eqs.\eqref{SWinnp}, the norm of $(f,f',g,g')$ as well as the inner products $\langle f|g\rangle$ and $\langle f'|g'\rangle$ are singular when $\lambda \approx 1$. To avoid a trivial vanishing of the correlation functions, we need to introduce the suitable scaling factor \cite{Summers:1987fn,Summers:1987squ}
\begin{equation} 
c_{\lambda} = \sqrt{\frac{1-\lambda^2}{1+\lambda^2}} \;, \label{scfact}
\end{equation}
and define Alice's and Bob's Weyl operators as 
\begin{equation} 
{\cal A}_{\hat f} = e^{i c_\lambda (a_{\hat f} + a^\dagger_{\hat f}) } \;, \qquad {\cal A}_{\hat f'} = e^{i c_\lambda (a_{\hat f'} + a^\dagger_{\hat f'}) } \;, \qquad {\rm Alice's\; Weyl \; operators} \;, \label{Alice}
\end{equation} 
and
\begin{equation} 
{\cal A}_{\hat g} = e^{i c_\lambda (a_{\hat g} + a^\dagger_{\hat g}) } \;, \qquad {\cal A}_{\hat g'} = e^{i c_\lambda (a_{\hat g'} + a^\dagger_{\hat g'}) } \;, \qquad {\rm Bob's\; Weyl \; operators} \;, \label{Bob}
\end{equation} 
with
\begin{eqnarray} 
{\hat f} &=& x f + y f'\;, \qquad  {\hat f'} = x'f + y'f'\;, \nonumber \\
{\hat g} &=& \colAFV{m g + n g'}\;, \qquad {\hat g'} = m'g + n'g'\;, \label{xy}
\end{eqnarray}
where $(x,y,m,n,x',y',m',n')$ are free real adjustable parameters, to be fine tuned in order to get sensible violations for the Bell-CHSH correlator $\langle 0 | {\cal C}| 0\rangle$, defined by 
\begin{equation} 
\langle 0 | {\cal C} | 0 \rangle = \langle 0 | ( {\cal A}_{\hat f} + {\cal A}_{\hat f'} ){\cal A}_{\hat g} + ( {\cal A}_{\hat f} - {\cal A}_{\hat f'} ){\cal A}_{\hat g'} |0\rangle 
\end{equation} 
From 
\begin{equation} 
 \langle 0 | {\cal A}_{\hat f} {\cal A}_{\hat g}| 0  \rangle = e^{-\frac{c^2_\lambda}{2} || {\hat f} + {\hat g}||^2} \;, \label{WW}
\end{equation}
we get, upon setting $\lambda=1$, 
\begin{equation} 
 \langle 0 | {\cal A}_{\hat f} {\cal A}_{\hat g}| 0  \rangle_{\lambda=1} = e^{-\frac{1}{2} (x^2+ y^2+m^2+n^2+ 2(xm - yn))} \;, \label{WW1}
\end{equation}
so that 
\begin{eqnarray} 
\langle 0 | {\cal C} | 0\rangle_{\lambda=1} & = & e^{-\frac{1}{2} (x^2+ y^2+m^2+n^2+ 2(xm - yn))}+ e^{-\frac{1}{2} (x'^2+ y'^2+m^2+n^2+ 2(x'm - \colAFV{y'n}))} \nonumber \\
& + & e^{-\frac{1}{2} (x^2+ y^2+m'^2+n'^2+ 2(xm' - yn'))} - e^{-\frac{1}{2} (x'^2+ y'^2+m'^2+n'^2+ 2(x'm' - y'n'))} \;. \label{WW2}
\end{eqnarray}
Maximizing this expression with respect to $(x,y,m,n,x',y',m',n')$, one finds
\begin{equation} 
\langle 0 | {\cal C} | 0\rangle_{\lambda=1} = 2.3244 \;, \label{vswW}
\end{equation}
which is in nice agreement with the value already given in Eq.\eqref{viol23}, obtained with a different set of vectors.

\subsubsection{Hunting Tsirelson's bound} 

Even if  the use of the Weyl operators, Eq.\eqref{weylpsi}, has enabled us to obtain a violation whose size is $\approx 2.3$, we have to remind that these operators are unitary. As such, we need to move to truly Hermitian bounded operators, a quite challenging task. \\\\In a first attempt, one might be tempted to employ operators which are well tested in Quantum Mechanics and try to generalize them to the case of Quantum Field Theory. Despite our efforts in this direction, we haven't been able to get too far. The example of the bounded Hermitian operator of eq.\eqref{par} illustrates this point in a clever way. \\\\Consider  in fact 
\begin{equation} 
A_{\hat f} = \Pi_{\hat f} =  \frac{1}{\pi} \int_{\mathbb{C}} d^2 \alpha \; e^{-|\alpha|^2}  
e^{\alpha a^\dagger_{\hat f} - \alpha^{*} a_{\hat f}} \;, \label{parsw}
\end{equation} 
and similar expressions for $A_{\hat f'}$, $A_{\hat g}$ and $A_{\hat g'}$, with ${\hat f},{\hat f'}, {\hat g}$ and ${\hat g'}$ given by Eq.\eqref{xy}. This operator is a bounded Hermitian operator, largely employed in Quantum Mechanics \cite{Banaszek:1998wcc}. After some algebra, for the Bell-CHSH correlation function we get 
\begin{eqnarray} 
\langle 0 | {\cal C} | 0\rangle^{\Pi}_{\lambda=1}  & = & \langle 0 |( \Pi_{\hat f}+\Pi_{\hat f'} )\Pi_{\hat g} + (\Pi_{\hat f}-\Pi_{\hat f'}) \Pi_{\hat g'} |0\rangle_{\lambda=1} \nonumber \\
& =& \frac{1}{1 + \frac{m^2-n^2}{2}}\left(  \frac{1}{1+ \frac{x^2+y^2}{2}}+   \frac{1}{1+ \frac{x'^2+y'^2}{2}}  \right) + \frac{1}{1 + \frac{m'^2-n'^2}{2}}\left(  \frac{1}{1+ \frac{x^2+y^2}{2}} -   \frac{1}{1+ \frac{x'^2+y'^2}{2}}  \right)  \;, \label{parxy}
\end{eqnarray}
whose maximization gives precisely $2$, {\it i.e.}
\begin{equation} 
\langle 0 | {\cal C} | 0\rangle^{\Pi}_{max} = 2 \;. \label{two}
\end{equation}
So, no Bell-CHSH violation. Other kinds of  Hermitian operators might be tested: $\sin(a_{\hat f} + a^\dagger_{\hat f})$, $ \cos(a_{\hat f} + a^\dagger_{\hat f})$, $\tanh(a_{\hat f} + a^\dagger_{\hat f})$, ${\rm sign}(a_{\hat f} + a^\dagger_{\hat f})$  \cite{Guimaraes:2026xnu}. So far, we haven't been able to find a clean sensible violation as found, for instance,  in the case of pure Weyl  unitaries.  \\\\Although the situation might not look encouraging, there is an important observation which permeates the whole reasoning of Summers-Werner's  work \cite{Summers:1987fn,Summers:1987squ,Summers:1987ze}, namely: violations of the Bell-CHSH inequality become accessible when the modular aspects of Quantum Field Theory are manifestly present in the construction of the Bell observables. Needless to say, the von Neumann algebra of the local operators of a  Quantum Field Theory is of type $III_1$, a fact that has deep and far-reaching consequences \cite{Summers:1987fn,Summers:1987squ,Summers:1987ze}. In practice, this means that the Bell operators have to encode, somehow, information about the spectrum of the modular operator $\delta$, a not easy goal. Here, we can provide a simple idea of how this mechanism might work by relying on a rather intuitive example. Consider in fact the operator 
\begin{equation} 
A_\epsilon(f) = \sqrt{\frac{\sqrt{2} (1-\lambda^2)}{(1+\lambda^2)}} (a_f + a^\dagger_f) e^{-\epsilon c^2_\lambda (a_f + a^\dagger_f)^2}  \;, \label{eps}
\end{equation} 
where $f$ stands for the Summers-Werner vector, Eq.\eqref{fsw}, $c_\lambda$ is the scaling factor of Eq.\eqref{scfact},  and $\epsilon$ is a non-vanishing, arbitrarily small, parameter. One notes that $ A_\epsilon(f) $ has a strong dependence from the spectral parameter $\lambda^2 \in [0,1]$. Essentially 
\begin{equation} 
A_\epsilon(f) = \sqrt{\frac{\sqrt{2} (1-\lambda^2)}{(1+\lambda^2)}} (a_f + a^\dagger_f)  + O(\epsilon) \;, \label{eps1}
\end{equation}
\colAFV{Strictly speaking, in the Bose case, the operator $(a_f + a^\dagger_f)$ is unbounded. Therefore Eq.\eqref{eps1} should not be read as a uniform operator-norm expansion. Rather, it is a shorthand for the leading matrix-element behavior on the scaled Summers-Werner vectors, the point of which is captured by the following expression}
\begin{equation} 
\langle 0| A_\epsilon(f) A_\epsilon(g) |0 \rangle = \sqrt{2} \;\frac{ 1-\lambda^2}{1+\lambda^2} \langle f | g \rangle + O(\epsilon) \;. \label{magic}
\end{equation}
Thus, for the Bell-CHSH inequality, we get 
\begin{eqnarray}
\langle 0 | {\cal C} | 0\rangle & = & \langle 0| (A_\epsilon(f) + A_\epsilon(f') )  A_\epsilon(g) + (A_\epsilon(f) - A_\epsilon(f') )  A_\epsilon(g') |0 \rangle \nonumber \\
& = & \sqrt{2} \;\frac{ 1-\lambda^2}{1+\lambda^2} \left( \langle f | g \rangle  +  \langle f' | g \rangle+  \langle f | g' \rangle - \langle f' | g' \rangle  \right) + O(\epsilon)  \;.  \label{mag2}
\end{eqnarray} 
Making now use of the inner products of Eq.\eqref{SWinnp}, \colAFV{one immediately realizes that the singular factor in $\langle f|g\rangle$ cancels the prefactor in Eq.\eqref{magic}, and} we end up with 
\begin{equation}
\langle 0 | {\cal C} | 0\rangle = 2 \sqrt{2} \frac{2 \lambda}{1+\lambda^2} + O(\epsilon) \;, \label{mag3}
\end{equation} 
so that, when $\lambda \approx 1$
\begin{equation} 
\langle 0 | {\cal C} | 0\rangle_{\lambda \approx 1}  \approx 2 \sqrt{2} + O(\epsilon)  \;, \label{mag4}
\end{equation}
we get arbitrarily close to Tsirelson's bound, exactly as in \cite{Summers:1987fn,Summers:1987squ,Summers:1987ze}. In other words, any operator fulfilling eq.\eqref{magic} will get close to $2\sqrt{2}$. \\\\At this stage, the difference with operators like that of eq.\eqref{par} or operators like $\sin$, $\cos$, $\tanh$, etc., becomes manifest. These operators produce essentially Gaussian correlations of the type 
\begin{equation} 
e^{- \mu^2 \langle f+g|f+g\rangle}  \;, \label{damp1}
\end{equation}
for some suitable constant factor $\mu$. This implies a fast decay as well as an eventually small, if not vanishing, violation. In the case of pure Weyl operators, the exponential factors survive, yielding the value $2.3$. However, in the case of expression \eqref{par} as well as of  other operators, an additional drastic damping takes place, due to the integral representation, {\it i.e.}
\begin{equation} 
\int dk \rho(k) \;e^{- k ^2 \mu^2 \langle f+g|f+g\rangle}  \;, \label{damp2}
\end{equation}
which gets even more suppressed. \\\\It is helpful here to elaborate more on the concrete realizability of Eq.\eqref{magic} in the bosonic case. To this end, we note the following observation:  Eq.\eqref{magic} is naturally fulfilled in the case of Fermi fields, see Sect. IV of \cite{Summers:1987squ}, where the proof of the saturation of Tsirelson's bound for free  Fermi fields has been given, see also \cite{Dudal:2023mij,Dudal:2024bmf}. The construction of the Summers-Werner vectors, Eqs.\eqref{fsw}, generalizes to the spinor case, implying that, when properly normalized, the smeared Hermitian combination 
\begin{equation} 
A_{\psi(f)} = \psi(f) + \psi(f)^\dagger \;, \label{spinf} 
\end{equation} 
where $\psi(f) = \int d^2x {\bar f}(x) \psi(x) $ is the smeared spinor field\footnote{In the Fermi case, the smearing procedure is achieved by a spinor test function $f$ \cite{Summers:1987squ}. The quantity ${\bar f}$ denotes the Dirac conjugation, ${\bar f}= f^\dagger \gamma^0$.  }, fulfills precisely condition \eqref{magic} \cite{Summers:1987squ}. Notice that, in the Fermi case, expression \eqref{spinf} is automatically bounded due to the algebra of the \colAFV{anti-commutation} canonical relations, see \colAFV{App.\eqref{appC}}. It seems thus that a possible way out in the bosonic case is to figure out an operator, built out from the Bose field, which behaves like a fermion. The concrete framework for this construction is provided by the bosonization technique of $1+1$ models \cite{Coleman:1985rnk}. Briefly, in such theories one relies on the notion of chirality, allowing to write down the so called vertex operators, namely 
\begin{equation} 
A_{vert}(h) = \int dx \;h(x)\; : e^{i \pi \phi_R(x) }: \;, \label{vert1}
\end{equation}
where $\phi_R$ is a right chiral boson field. Such operators enjoy precisely property \eqref{magic}, since $A_{vert}(h)$ behaves in fact like a Fermi field. Bosonization seems thus to give the natural and concrete setup to address the issue of the violation of the Bell-CHSH inequality in the Bose case. We shall report on this fascinating topic in a more extended and detailed work. \\\\For the benefit of the reader, in App.\eqref{appC}, one finds a detailed summary of the Summers-Werner argument leading to maximal violation for free Fermi fields.

\section{Conclusion}\label{sect5}

In this work, the deep relationship between the Bell-CHSH inequality and the modular theory of Tomita-Takesaki in relativistic Quantum Field Theory has been addressed. \\\\The aim was that of devising a framework enabling us to employ the modular theory to investigate whether violations of the Bell-CHSH inequality might occur. \\\\This has been done by elaborating on the following issues: 
\begin{itemize} 

\item after providing a short background on the Tomita-Takesaki modular theory, the modular operators $(s,\delta,j)$ have been employed to give examples of vectors of the 1-particle Hilbert space localized in the right and left wedges $(W_R,W_L)$. In particular, the construction of the four vectors $(f,f',g,g')$ employed by Summers-Werner in their proof of maximal violation has been addressed in detail. 

\item the not easy issue of the choice of a suitable set of Bell operators able to violate the Bell-CHSH inequality in the vacuum state has been faced. We have pinpointed, through the example of the $\Pi_f$ operator, that a mere generalization to Quantum Field Theory of the bounded Hermitian operators employed in Quantum Mechanics does not suffice to saturate Tsirelson's bound. 

\item the necessity of building operators capable of feeling the spectrum of the modular operator $\delta$ has been pointed out and illustrated with a simple intuitive example. \colAFV{This is a rather deep point that underlines the qualitative difference from ordinary quantum mechanics: in our case the local algebras are of type $\mathrm{III}_1$, which means that no normal pure state exists on either wedge algebra.}

\end{itemize}

\noindent The explicit construction of nice operators allowing to saturate Tsirelson's bound remains a challenging matter. Moreover, we highlighted that the so called vertex operators of bosonization theory might provide a quite concrete possibility, potentially opening connections with holography and other areas.

\section*{Acknowledgments}
The authors would like to thank the Brazilian agencies CNPq, CAPES and FAPERJ for financial support.  S. P.~Sorella, I.~Roditi, and M. S.~Guimaraes are CNPq researchers under contracts 302991/2024-7, 311876/2021-8, and 309793/2023-8, respectively. Prof. Ricardo Correa da Silva is gratefully acknowledged for fruitful discussion. 
%\end{acknowledgments}

\appendix

\section{Canonical quantization }\label{appA}

\subsection{The real massive scalar field in $(1+1)$ Minkowski spacetime}

We consider a free real scalar field of mass $m$ in $(1+1)$-dimensional Minkowski spacetime. Its quantized form, expressed via plane-wave decomposition, reads
	\begin{equation}\label{qf}
		\varphi(t,{x}) = \int \! \frac{d k}{(2 \pi)} \frac{1}{2 \omega_k} \left( e^{-ik_\mu x^\mu} a_k + e^{ik_\mu x^\mu} a^{\dagger}_k \right) \;, \qquad k_\mu x^\mu = \omega_k t - {k}  { x} \;,
	\end{equation}
where $\omega_k = \sqrt{{k}^2 + m^2}$ and the annihilation and creation operators satisfy the canonical commutation relations
	\begin{align}\label{eq:CCR}
		[a_k, a^{\dagger}_q] &= (2\pi) \, 2\omega_k \, \delta(k - q), \\ \nonumber 
		[a_k, a_q] &= [a^{\dagger}_k, a^{\dagger}_q] = 0. 
	\end{align}

In a rigorous QFT framework, the scalar field is treated as an operator-valued distribution \cite{Haag:1992hx}, and thus must be smeared out with smooth, compactly supported test functions to yield well-defined operators in the Hilbert space. Given a smooth, compactly supported, test function $f(\vb*{x}) \in C_0^\infty(\mathbb{R}^{2})$,  the corresponding smeared field operator is defined as
\begin{align} 
		\varphi(f) = \int \! d^2x \; \varphi(t,x) f(t,x) \;.
	\end{align}
The vacuum expectation value of two smeared field operators, given by  the two-point Wightman function, defines the Lorentz invariant inner product  $\langle f \vert g \rangle$:
\begin{align} \label{Inn}
		\langle f \vert g \rangle &= \langle 0 \vert \varphi(f) \varphi(g) \vert 0 \rangle = i \Delta_{\rm PJ}(f,g) +  H(f,g),
	\end{align}
where $f, g \in C_0^\infty$, and $\Delta_{\text{PJ}}(f,g)$ and $H(f,g)$ denote the smeared Pauli–Jordan and Hadamard distributions, respectively. These are defined by
\begin{align}
		\Delta_{\rm PJ}(f,g) &=  \int \! d^2x d^2y f(x) \Delta_{\rm PJ}(x-y) g(y) \;,  \nonumber \\
		H(f,g) &=  \int \! d^2x d^2y f(x) H(x-y) g(y)\;. \label{mint}
	\end{align}
with distributional kernels given explicitly by
\begin{eqnarray} 
\Delta_{\rm PJ}(x) & =& - \int \frac{dp}{(2\pi)} \frac{1}{2\omega_k} \sin\left(p_\mu (x-x')^\mu \right)  \;, \nonumber \\
H(x) & = & \int \frac{dp}{(2\pi)} \frac{1}{2\omega_k} \cos\left( p_\mu (x-x')^\mu \right)  \;. \label{PJH}
\end{eqnarray}
Both  $\Delta_{\rm PJ}(x)$  and $H(x)$ are Lorentz-invariant. The Pauli-Jordan distribution  $\Delta_{\rm PJ}(x)$  encodes the principle of causality, vanishing outside the light cone. Moreover, it is odd under $x \rightarrow -x$, whereas the Hadamard function $H(x)$ is even. The commutator of the field operators is thus $\left[\varphi(f), \varphi(g)\right] = i \Delta_{\rm PJ}(f,g)$, ensuring that  $\left[\phi(f), \phi(g)\right] = 0,$ whenever the supports of $f$ and $g$ are spacelike separated. This compactly encodes the principle of micro-causality in relativistic field theory. \\\\In momentum space, we have  
\begin{equation} 
\langle f \vert g \rangle = \int \frac{d^2p}{(2 \pi)^2} \; (2\pi) \theta(p^0)\delta(p^2 -m^2) f^{*}(p_0,p) g(p_0,p) \;, \label{fin}
\end{equation} 
 where $(f(p_0,p),g(p_0,p))$ stand for the Fourier transform of $(f(x),g(x))$: 
 \begin{equation} 
 f(p_0,p) = \int d^2x \; e^{i p_\mu x^\mu} f(x) \;. \label{ft}
 \end{equation}

 \section{The Reeh-Schlieder theorem for the 1-particle Hilbert space}\label{appB}
 
 This Appendix is devoted to a sketchy proof of the so called Reeh-Schlieder theorem for the 1-particle Hilbert space, see 
\cite{Guido:2008jk} and refs. therein. We consider a free real massive scalar field $\varphi(f)$, $f \in {\cal S}({\mathbb{R}}^{1+d})$ with support contained in an open non empty  connected region ${\cal O}$ of the \colAFV{Minkowski} spacetime $M^{1+d}, d=1,2,3$. \\\\The 1-particle Hilbert space of the theory is 
\begin{equation} 
{\cal H}_{1-p} = L^2(d\mu_p, H_m) \;, \qquad d\mu_p = \frac{d^dp}{(2\pi)^d}\frac{1}{2 \omega_p} \;, \qquad H_m =\{ p^\mu,\; p^2=m^2, \omega_p>0\} \;. \label{1pB}
\end{equation}
One considers the embedding from the space of the test functions $\{ f\}$ to the Hilbert space ${\cal H}_{1-p}$ specified by 
\begin{equation} 
\psi_f(p) = \int d^{1+d}x \; e^{-i (\omega_p t - {\vec p}\cdot {\vec x})} f(t,x) \;, \qquad supp(f) \in {\cal O} \;, \qquad p^2=m^2 \;. \label{emb}
\end{equation}
The subspace $K({\cal O})$ of $L^2(d\mu_p, H_m)$ is defined as the closure of the real subspace 
\begin{equation} 
K({\cal O}) = \{ \psi_f(p) = \int d^{1+d}x \; e^{-i (\omega_p t - {\vec p}\cdot {\vec x})} f(t,x) \;, \;\;\;\; supp(f) \in {\cal O} \; \} \;. \label{KO}
\end{equation}
The real subspace $K({\cal O})$ can be proven to be a standard subspace for ${\cal H}_{1-p}$ meaning, in particular, that $K({\cal O})+iK({\cal O})$ is dense in ${\cal H}_{1-p}$. A sketchy proof of this important statement can be provided as follows. \\\\Suppose that there exists a $\psi(p) \in {\cal H}_{1-p}$ orthogonal to any element of $K({\cal O})+iK({\cal O})$, {\it i.e.} 
\begin{equation} 
\langle \psi | \psi_g\rangle = 0 \;\;\;\forall  g \;, \;\;\; supp(g) \in {\cal O} \;, \label{abs1}
\end{equation} 
namely 
\begin{equation}
\int d^dp\; \psi(p)^* \psi(p)_g =\int d^dp \;\psi(p)^* \int d^{1+d}x \; e^{- ipx} g(t,x)   = 0  \;\;\;\forall  g \;, \;\;\; supp(g) \in {\cal O} \;. \label{abs2}
\end{equation}
Equation \eqref{abs2} can be rewritten as 
\begin{equation} 
\int d^{1+d}x\; F(t,x) g(t,x) = 0  \;\;\;\forall  g \;, \;\;\; supp(g) \in {\cal O} \;, \label{abs3}
\end{equation}
where $F(t,x)$ is given by 
\begin{equation} 
F(t,x) = \int d^d p\; \psi(p)^* e^{-i (\omega_p t - {\vec p}\cdot {\vec x})} \;. \label{abs4}
\end{equation} 
Now one observes that $F(t,x)$ exhibits analytic continuation in the complex time variable $ z=t-iy, y\ge0$, as it follows from the presence of the damping factor $ e^{-y \omega_p}$ in 
\begin{equation} 
F(z,x) = F(t-iy,x) = \int d^d p\; \psi(p)^* e^{-i (\omega_p t - {\vec p}\cdot {\vec x})} e^{-y \omega_p} \;. \label{abs5} 
\end{equation} 
Thus, $F(z,x)$ is an analytic function in the region $(t-iy,x)$ which, moreover, due to \eqref{abs3} vanishes in the open region $(t,x) \in {\cal O}$. Therefore, from the fundamental theorem on analytic functions, $F$ vanishes identically, implying, in turn, that $\psi(p)=0$, {\it i.e.} no non-trivial orthogonal vectors exist, showing thus the desired property.

\section{Saturation of the Tsirelson bound for free Fermi fields}\label{appC}	

This Appendix is devoted to provide a detailed summary of the saturation of the Tsirelson bound in the case of free Fermi fields. As already underlined, the spinor case allows for a very clean proof of the maximal violation of the Bell-CHSH inequality in the vacuum state, due precisely to property eq.\eqref{magic}. As explicit example, we shall discuss the case of a massless Majorana field in $1+1$, namely 
\begin{equation} 
S = \int d^2x \;\frac{i}{2} {\bar \psi} \gamma^\mu \partial_\mu \psi \;, \label{Mact}
\end{equation}
with 
\begin{equation} 
\gamma^0=\begin{pmatrix}
0 & 1 \\
1 & 0
\end{pmatrix} \;, \qquad
\gamma^1=\begin{pmatrix}
0 & 1 \\
-1 & 0
\end{pmatrix}\;, \qquad g_{\mu\nu}=diag(1,-1) \;. \label{conv}
\end{equation}
The Majorana condition 
\begin{equation} 
\psi^{C} = \psi \;, \qquad \psi^{C} = C {\bar \psi}^T \;, \qquad C=\gamma^1 \;, \label{mj1}
\end{equation}
gives 
\begin{equation} 
\psi = \begin{pmatrix}
h  \\
i \varphi 
\end{pmatrix} \;, \label{mj2}
\end{equation}
with $(h,\varphi)$ real. Moreover, from the equations of motion, we get 
\begin{equation} 
(\partial_t + \partial_x)\varphi =0 \;, \qquad (\partial_t - \partial_x) h = 0 \;, \label{mj3}
\end{equation}
so that, making use of the light cone coordinates $(x_+=x+t, x_-=x-t)$, it follows that 
\begin{equation} 
h = h(x_+) \;, \qquad \varphi = \varphi(x_-) \;. \label{lrm}
\end{equation}
The real components $(h,\varphi)$ are referred to as the left $(L)$ and right $(R)$ movers. For the plane wave expansion, we have 
\begin{eqnarray} 
h(x_+) & = & \int_{-\infty}^0 \frac{dk}{2 \pi} \frac{1}{\colAFV{\sqrt{2|k|}}} \left( a_k e^{-i kx_+} + a^\dagger_k e^{ikx_+} \right) \;, \nonumber \\
\varphi(x_-) & = & \int_0^\infty \frac{dk}{2 \pi} \frac{1}{\sqrt{2k}} \left( b_k e^{-i kx_-} + b^\dagger_k e^{ikx_-} \right) \;, 
\end{eqnarray}
with 
\begin{eqnarray} 
\{ a_k, a^\dagger_q \} &=& (2\pi) \;2 |k| \;\delta(k-q) \;, \qquad \{ a_k, a_q \} =0 \;, \nonumber \\
\{ b_k, b^\dagger_q \} &=& (2\pi) \;2 |k| \;\delta(k-q) \;, \qquad \{ b_k, b_q \} =0 \;, \nonumber \\
\{ a_k, b_q \} &=& 0  \;, \qquad \{ a_k, b^\dagger_q \} =0 \;. \label{car}
\end{eqnarray}
Having two real components, the smearing of the Majorana spinor requires only two smooth test functions: $(f_1(x_+),f_2(x_-)) \in {\cal C}_0^\infty({\mathbb{R}})$: 
\begin{equation} 
\psi(f) = \int_{-\infty}^{\infty} dx_+ \;f_1(x_+) h(x_+) + \int_{-\infty}^{\infty} dx_- \; f_2(x_-) \varphi(x_-) \;. \label{smsp}
\end{equation}
Setting 
\begin{equation} 
f_1(k) = \int_{-\infty}^{\infty} dx_+ \; e^{i kx_+} f_1(x_+)\;, \qquad f_2(k) = \int_{-\infty}^{\infty} dx_- \; e^{i kx_-} f_2(x_-)\;, \label{cft}
\end{equation}
one has 
\begin{equation} 
\psi(f) =(a_f + a^\dagger_f + b_f + b^\dagger_f) \;, \label{smspsp}
\end{equation}
where $(a_f,b_f)$ are the smeared annihilation operators, {\it i.e.} 
\begin{equation} 
a_f= \int_{-\infty}^0  \frac{dk}{2 \pi} \frac{1}{\colAFV{\sqrt{2|k|}}} \;f^{*}_1(k) a_k \;, \qquad b_f= \int_0^\infty  \frac{dk}{2 \pi} \frac{1}{\sqrt{2k}} \;f^{*}_2(k) b_k \;. \label{abnnop}
\end{equation}
The smeared Majorana field is, by \colAFV{construction}, Hermitian, $\psi(f) = \psi^{\dagger}(f)$. Moreover, unlike the Bose case, the anti-commutation relations \eqref{car} ensure that $\psi(f)$ is a bounded operator. For the inner product $\langle f|g\rangle$, we have now 
\begin{equation} 
\langle 0| \psi(f) \psi(g) |0\rangle = \langle f|g\rangle = \int_{-\infty}^0 \frac{dk}{2 \pi} f^{*}_1(k) g_1(k) + \int_0^\infty \frac{dk}{2 \pi} f^{*}_2(k) g_2(k)  \;, \label{fgMj}
\end{equation}
from which it follows that the 1-particle Hilbert space ${\cal H}$ is given by the direct sum of the Hilbert spaces corresponding to the right and left chirality: 
\begin{equation} 
{\cal H} = {\cal H}_+ \oplus {\cal H}_- \;, \qquad {\cal H}_+= L^2\left((-\infty,0),\frac{dk}{2 \pi}\right) \;, \qquad {\cal H}_-= L^2\left((0,\infty),\frac{dk}{2 \pi}\right) \;. \label{Hpm}
\end{equation} 
In order to introduce the standard subspaces \cite{Brunetti:1992zf,Borchers:1998ye}, it is helpful to notice that, \colAFV{in terms of} the light cone coordinates $(x_+,x_-)$, the right wedge $W_R$ can be described as 
\begin{equation} 
W_R =\{ x_+>0\; \cap \;x_-<0 \}  \;. \label{WRpm}
\end{equation} 
The left movers see the region $x_+>0$, while the right movers the region $x_-<0$. One introduces the two real subspaces $K_+(W_R)$ and $K_-(W_R)$ defined as 
\begin{eqnarray} 
K_+(W_R) & = & \{ P_+ f_+(k)\;, f_+(k) = \int_0^\infty dx_+ e^{i k x_+} f_+(x_+)\;, f_+\in{\cal C}_0^\infty(0,\infty) \;, P_+f_+(k) = f_+(k)|_{k>0} \} \nonumber \\
K_-(W_R) & = & \{ P_- f_-(k)\;, f_-(k) = \int_{-\infty}^0 dx_- e^{i k x_-} f_-(x_-)\;, f_-\in{\cal C}_0^\infty(-\infty,0) \;, P_-f_-(k) = f_-(k)|_{k<0} \} \;. \label{stst}
\end{eqnarray}
The real standard subspace $K(W_R)$ is thus defined as 
\begin{equation} 
K(W_R) = K_+(W_R) \oplus K_-(W_R) \;, \label{stMj}
\end{equation}
with
\begin{itemize} 
\item 
\begin{equation} 
K(W_R) \cap iK(W_R) =\{  0 \} \;, \label{Mjempty} 
\end{equation}

\item 
\begin{equation} 
K(W_R) + i K(W_R) \;\;\; \mathrm{ is \; dense  \,  \; in \; the \; Hilbert \; space \; {\cal H}} \;.\label{denseMJ}
\end{equation}

\end{itemize}
Concerning the analytic continuation, it is easy to check out that vectors belonging to $K_+(W_R)$ admit an analytic continuation in the upper half plane $Im(k)>0$, while vectors of $K_-(W_R)$ can be analytically continued in the lower half plane $Im(k)<0$. \\\\Let us move now to the modular theory. Introducing the rapidity variable 
\begin{equation} 
k = e^{\theta} \;, \qquad dk = d\theta e^{\theta} \;, \label{rapMj}
\end{equation} 
the inner product $\langle f| g\rangle$ can be rewritten as 
\begin{equation} 
\langle f | g\rangle = \int_{-\infty}^{\infty} \frac{d \theta}{2\pi}  \left( e^{-\theta} f^{*}_1(\theta) g_1(\theta) + e^{\theta} f_2^{*}(\theta) g_2(\theta) \right) \;. \label{iprmj}
\end{equation}
A boost transformation with parameter $s$ amounts to a shift in rapidity. More precisely 
\begin{eqnarray} 
\theta & \rightarrow & \theta +s \;, \qquad { R}-sector \;, \nonumber \\
\theta & \rightarrow & \theta -s \;, \qquad { L}-sector \;. \label{bLR}
\end{eqnarray}
From the Bisognano-Wichmann results, it follows that the modular operator $\delta$ is related to the boosts generator by 
\begin{equation} 
\delta = e^{-2\pi K} \;, \qquad K=K_R + K_L \;, \label{BWMj}
\end{equation} 
where $(K_L,K_R)$ act, respectively, on the $R$ and $L$ sectors: 
\begin{equation} 
K_R= - i \frac{\partial}{\partial \theta} \;, \qquad K_L=  i \frac{\partial}{\partial \theta} \;. \label{BWactRL}
\end{equation}
Therefore, for a two component spinor 
\begin{equation} 
\xi(\theta) = \begin{pmatrix}
\xi_+(\theta)  \\
\xi_-(\theta)
\end{pmatrix} \;, \label{ximt}
\end{equation}
we have 
\begin{equation} 
\delta^{1/2} \xi_+(\theta) = \xi_+(\theta+i \pi) \;, \qquad \delta^{1/2} \xi_-(\theta) = \xi_-(\theta-i \pi) \;, \label{dximj}
\end{equation}
Again, from \cite{Bisognano:1975ih}, one learns that the modular conjugation $j$ is given by 
\begin{equation} 
j = R_3(\pi) (CPT) \;, \label{BWJMj}
\end{equation} 
where $R_3(\pi)$ stands for a rotation of $\pi$ around the $x$-axis, {\it i.e.}
\begin{equation} 
R_3(\pi) = e^{-\frac{i\pi}{2} \sigma_1}  \;, \label{Rfac}
\end{equation} 
where $\sigma_1$ is the Pauli matrix along the $x$-direction. Thus, up to an irrelevant global sign, one has 
\begin{equation} 
j=\begin{pmatrix}
0 & i \\
i & 0
\end{pmatrix} \; (CPT) \;,  \label{JSMj}
\end{equation}
{\it i.e.}
\begin{equation} 
j \xi_+(\theta) = i {\xi_-}^{*}(\theta)  \;, \qquad j \xi_-(\theta) = i {\xi_+}^{*}(\theta) \;, \label{jexc}
\end{equation}
from which one sees that $j$ exchanges the two chiral sectors $(R,L)$. From the knowledge of $\delta$ and $j$, one introduces the anti-linear Tomita-Takesaki operator $s$, defined by 
\begin{equation} 
s = j \delta^{1/2} \;, \label{TTMj}
\end{equation}
with 
\begin{equation} 
s^2 =1 \;, \qquad  j \delta^{1/2} j = \delta^{-1/2} \;, \qquad j^2=1 \;. \label{MjJJ}
\end{equation}
A two component spinor $\xi(\theta)$ is said to be localized in the right wedge $W_R$ when 
\begin{equation} 
s \;\xi(\theta) = \xi(\theta) \;. \label{Rspl}
\end{equation}
On the other hand, ${\hat \xi}(\theta)$ is localized in $W_L$ when\footnote{ The factor $i$ in expression \eqref{Lloc} is due to the so called twisted duality of Fermi fields \cite{Foit:1983pbi,Mund:2005cv}: $K(W_R)'= iK(W'_R)$.} 
\begin{equation} 
{\hat \xi}= i \eta \;, \qquad  s^\dagger {\eta }(\theta) = {\eta}(\theta) \;, \qquad s^\dagger = j \delta^{-1/2} \;. \label{Lloc}
\end{equation}
Making use of expressions \eqref{dximj},\eqref{jexc}, it turns out that condition \eqref{Rspl} reads
\begin{equation} 
s \;\xi(\theta) = \xi(\theta) \rightarrow  \xi_+(\theta) = i (\xi_-(\theta-i \pi))^{*} \;, \qquad  \xi_-(\theta) = i (\xi_+(\theta+i \pi))^{*} \;, \label{mjkms}
\end{equation}
Of course, a spinor of the kind 
\begin{equation} 
\xi = (1+s) {\cal F} \;, \label{projmj}
\end{equation}
fulfills the condition $s\xi =\xi$, being $W_R$-localized. 

\subsubsection{The violation of the Bell-CHSH inequality}\label{MJTs}

Having discussed the modular operators, we can face now the Bell-CHSH inequality. Following \cite{Summers:1987fn,Summers:1987squ,Summers:1987ze}, the first step is that of constructing a dichotomic operator. In the Fermi case, due to the anti-commutation relations, this task can be achieved by making use of the spinor field $\psi(f)$ itself: 
\begin{equation} 
A_f = \sqrt{\frac{1}{||f||^2}} \;\psi(f) \;, \qquad A^2_f = 1 \;. \label{dicMj}
\end{equation}
Therefore, for the Bell-CHSH correlation function in the vacuum, we get 
\begin{equation} 
\langle 0| {\cal C} |0\rangle = i \langle 0| (A_f + A_{f'})A_g + (A_f-A_{f'})A_{g'}|0\rangle = i \left( \frac{\langle f|g\rangle}{\sqrt{||f||^2 ||g||^2}} + \frac{\langle f'|g\rangle}{\sqrt{||f'||^2 ||g||^2}} + \frac{\langle f|g'\rangle}{\sqrt{||f||^2 ||g'||^2}} -\frac{\langle f'|g'\rangle}{\sqrt{||f'||^2 ||g||^2}} \right)   \label{BellMj}
\end{equation}
with $(f,f')$ and $(g,g')$ being, respectively,  $W_R$ and $W_L$ localized. Also, the factor $i$ in expression \eqref{BellMj} is due to the anti-commuting nature of $A_f$, {\it i.e.} 
\begin{equation}
(A_f A_g)^\dagger = A^\dagger_g A^\dagger_f = A_g A_f =- A_f A_g  \;. \label{ifact}
\end{equation}
It remains to choose the four vectors $(f,f',g,g')$. As outlined in \cite{Summers:1987fn,Summers:1987squ,Summers:1987ze}, this is done in two steps. First, one picks up a normalized $\Phi$ belonging to the spectral interval of the modular operator $\delta$ corresponding to $[\lambda^2-\epsilon, \lambda^2+\epsilon]$, with $\lambda^2 \in [0,1]$. Following \cite{Summers:1987fn,Summers:1987squ,Summers:1987ze}, one introduces the four vectors 
\begin{eqnarray} 
{\hat f} &=& \frac{1}{\sqrt{1-\mu_+}}(1+s)\Phi \;, \qquad {\hat f'} = \frac{1}{\sqrt{1-\mu_+}}(1+s) i \Phi \;, \nonumber \\
{\hat g} &=&i \frac{1}{\sqrt{\mu_- -1}}(1+s^\dagger)\Phi \;, \qquad {\hat g'} =i \frac{1}{\sqrt{\mu_- - 1}}(1+s^\dagger) i \Phi \;, \label{SWMj}
\end{eqnarray}
with 
\begin{equation} 
\mu_{+} = \langle \Phi | \delta | \Phi\rangle \underset{\epsilon \rightarrow 0}\rightarrow \lambda^2 \;, \qquad \mu_{-} = \langle \Phi | \delta^{-1} | \Phi\rangle \underset{\epsilon \rightarrow 0}\rightarrow \lambda^{-2} \;. \label{epslMj}
\end{equation}
At this stage, the only difference of expressions \eqref{SWMj} with respect to eqs.\eqref{fsw} is the extra factor $i$ appearing in $({\hat g},{\hat g'})$. The vectors $({\hat f}, {\hat f'}, {\hat g},{\hat g'})$ turn out to fulfill the conditions 
\begin{eqnarray} 
\langle {\hat f}|{\hat f'}\rangle & = & \langle {\hat g}|{\hat g'}\rangle = i  \;, \nonumber \\
||{\hat f}||^2 & = & ||{\hat f'}||^2 =||{\hat g}||^2 =||{\hat g'}||^2 = \frac{1+\lambda^2}{1-\lambda^2} \;, \nonumber \\
\langle {\hat f}|{\hat g'}\rangle & = &\langle {\hat f'}|{\hat g }\rangle  = 0 \;, \nonumber \\
\langle {\hat f}|{\hat g}\rangle & = & -\langle {\hat f'}|{\hat g'}\rangle  = \frac{2i\lambda}{1-\lambda^2} \;. \label{mjw1}
\end{eqnarray}
Thus,  the final form of $(f,f',g,g')$ is obtained by setting 
\begin{eqnarray} 
f &=& \sqrt{\frac{1-\lambda^2}{1+\lambda^2}} {\hat f} \;, \qquad ||f||^2=1 \;, \nonumber \\
f' &=& \sqrt{\frac{1-\lambda^2}{1+\lambda^2}} {\hat f'} \;, \qquad \colAFV{||f'||^2=1} \;, \nonumber \\
g &=&- \sqrt{\frac{1-\lambda^2}{1+\lambda^2}} \;\frac{({\hat g}- {\hat g'})}{\sqrt{2}} \;, \qquad ||g||^2=1 \;, \nonumber \\
g' &=&- \sqrt{\frac{1-\lambda^2}{1+\lambda^2}} \;\frac{({\hat g}+ {\hat g'})}{\sqrt{2}} \;, \qquad \colAFV{||g'||^2=1} \;. \label{mjw2}
\end{eqnarray} 
Plugging expressions \eqref{mjw2} into the Bell-CHSH correlator and using eqs.\eqref{mjw1} one gets 
\begin{eqnarray} 
\langle 0| {\cal C} |0\rangle &=& \frac{-i}{\sqrt{2}} \frac{1-\lambda^2}{1+\lambda^2}\left(\langle {\hat f}|{\hat g}-{\hat g'} \rangle  + \langle {\hat f'}|{\hat g}-{\hat g'} \rangle  + \langle {\hat f}|{\hat g}+{\hat g'}\rangle - \langle {\hat f'}|{\hat g}+{\hat g'} \rangle \right)
 \nonumber \\
&=&  \frac{-i}{\sqrt{2}} \frac{1-\lambda^2}{1+\lambda^2} \frac{8i \lambda}{1-\lambda^2} = 2\sqrt{2} \frac{2 \lambda}{1+\lambda^2}  \underset{\lambda \rightarrow 1} \rightarrow 2 \sqrt{2} \;, \label{mjw3}
\end{eqnarray}
showing the saturation of Tsirelson's bound when $\lambda \approx 1$. \\\\As final remark, one should notice that the normalization of the final form of the vectors $(f,f',g,g')$, as expressed by eqs.\eqref{mjw2}, exhibits precisely the factor $\sqrt{1-\lambda^2}$ of eq.\eqref{eps1}.

%%%%%%%%%%%%%%%%%%%%%%%%%%%%%%%%%%%%%%%%%%%%%%%%%%%%%%%%%%%%%%%%%%%%%%%%%%%%%%%%%%%%%%%%%%%%%%%%%%%%%%%%%%


\begin{thebibliography}{99}

%\cite{Brunetti:2002nt,Borchers:2000pv,Schroer:2014kya}

%\cite{Brunetti:2002nt}
\bibitem{Brunetti:2002nt}
R.~Brunetti, D.~Guido and R.~Longo,
%``Modular localization and Wigner particles,''
Rev. Math. Phys. \textbf{14} (2002), 759-786
doi:10.1142/S0129055X02001387
[arXiv:math-ph/0203021 [math-ph]].
%130 citations counted in INSPIRE as of 09 Feb 2026


%\cite{Borchers:2000pv}
\bibitem{Borchers:2000pv}
H.~J.~Borchers,
%``On revolutionizing quantum field theory with Tomita's modular theory,''
J. Math. Phys. \textbf{41} (2000), 3604-3673
doi:10.1063/1.533323
%191 citations counted in INSPIRE as of 09 Feb 20264

%\cite{Schroer:2014kya}
\bibitem{Schroer:2014kya}
B.~Schroer,
%``The Ongoing Impact of Modular Localization on Particle Theory,''
SIGMA \textbf{10} (2014), 085
doi:10.3842/SIGMA.2014.085
[arXiv:1407.2124 [math-ph]].
%7 citations counted in INSPIRE as of 09 Feb 2026



%\cite{Takesaki:1970aki}
\bibitem{Takesaki:1970aki}
M.~Takesaki, {\it Tomita's Theory of Modular Hilbert Algebras and its Applications},
Springer-Verlag, 1970, 
%``Tomita's Theory of Modular Hilbert Algebras and its Applications,''
doi:10.1007/bfb0065832
%59 citations counted in INSPIRE as of 06 Feb 2026

%\cite{Bratteli:1979tw}
\bibitem{Bratteli:1979tw}
O.~Bratteli and D.~W.~Robinson,
`{\it Operator  Algebras and  Quantum Statistical Mechanics, 1.} , Springer-Verlag (1987)
%8 citations counted in INSPIRE as of 06 Feb 2026


%\cite{Summers:2003tf}
\bibitem{Summers:2003tf}
S.~J.~Summers,
%``Tomita-Takesaki modular theory,''
[arXiv:math-ph/0511034 [math-ph]].
%36 citations counted in INSPIRE as of 06 Feb 2026

%\cite{Guido:2008jk}
\bibitem{Guido:2008jk}
D.~Guido,
%``Modular theory for the von Neumann algebras of Local Quantum Physics,''
Contemp. Math. \textbf{534} (2011), 97-120
[arXiv:0812.1511 [math.OA]].
%19 citations counted in INSPIRE as of 06 Feb 2026

%\cite{Witten:2018zxz}
\bibitem{Witten:2018zxz}
E.~Witten,
%``APS Medal for Exceptional Achievement in Research: Invited article on entanglement properties of quantum field theory,''
Rev. Mod. Phys. \textbf{90} (2018) no.4, 045003
doi:10.1103/RevModPhys.90.045003
[arXiv:1803.04993 [hep-th]].
%674 citations counted in INSPIRE as of 06 Feb 2026


%\cite{Bisognano:1975ih}
\bibitem{Bisognano:1975ih}
J.~J.~Bisognano and E.~H.~Wichmann,
%``On the Duality Condition for a Hermitian Scalar Field,''
J. Math. Phys. \textbf{16} (1975), 985-1007
doi:10.1063/1.522605
%556 citations counted in INSPIRE as of 06 Feb 2026



%\cite{Bell:1964kc}
\bibitem{Bell:1964kc}
J.~S.~Bell,
%``On the Einstein-Podolsky-Rosen paradox,''
Physics Physique Fizika \textbf{1} (1964), 195-200
doi:10.1103/PhysicsPhysiqueFizika.1.195
%3931 citations counted in INSPIRE as of 06 Feb 2026

%\cite{Clauser:1969ny}
\bibitem{Clauser:1969ny}
J.~F.~Clauser, M.~A.~Horne, A.~Shimony and R.~A.~Holt,
%``Proposed experiment to test local hidden variable theories,''
Phys. Rev. Lett. \textbf{23} (1969), 880-884
doi:10.1103/PhysRevLett.23.880
%5430 citations counted in INSPIRE as of 06 Feb 2026



%\cite{Summers:1987fn}
\bibitem{Summers:1987fn}
S.~J.~Summers and R.~Werner,
%``BellInequalities and Quantum Field Theory. 1. General Setting,''
J. Math. Phys. \textbf{28} (1987), 2440-2447
doi:10.1063/1.527733
%216 citations counted in INSPIRE as of 06 Feb 2026

%\cite{Summers:1987squ}
\bibitem{Summers:1987squ}
S.~J.~Summers and R.~Werner,
%``Bell{\textquoteright}s inequalities and quantum field theory. II. Bell{\textquoteright}s inequalities are maximally violated in the vacuum,''
J. Math. Phys. \textbf{28} (1987) no.10, 2448-2456
doi:10.1063/1.527734
%116 citations counted in INSPIRE as of 06 Feb 2026

%\cite{Summers:1987ze}
\bibitem{Summers:1987ze}
S.~J.~Summers and R.~Werner,
%``Maximal Violation of Bell's Inequalities Is Generic in Quantum Field Theory,''
Commun. Math. Phys. \textbf{110} (1987), 247-259
doi:10.1007/BF01207366
%149 citations counted in INSPIRE as of 06 Feb 2026



%\cite{Guimaraes:2024byw}
\bibitem{Guimaraes:2024byw}
M.~S.~Guimaraes, I.~Roditi and S.~P.~Sorella,
%``Introduction to Bell{\textquoteright}s Inequality in Quantum Mechanics {\textdagger},''
Universe \textbf{10} (2024) no.10, 396
doi:10.3390/universe10100396
[arXiv:2409.07597 [quant-ph]].
%8 citations counted in INSPIRE as of 06 Feb 2026

%\cite{Guimaraes:2024mmp}
\bibitem{Guimaraes:2024mmp}
M.~S.~Guimaraes, I.~Roditi and S.~P.~Sorella,
%``Bell{\textquoteright}s inequality in relativistic Quantum Field Theory,''
Rev. Phys. \textbf{13} (2025), 100121
doi:10.1016/j.revip.2025.100121
[arXiv:2410.19101 [quant-ph]].
%8 citations counted in INSPIRE as of 06 Feb 2026



%\cite{Dudal:2023mij,Dudal:2024bmf,Dudal:2026eil}


%\cite{Dudal:2023mij}
\bibitem{Dudal:2023mij}
D.~Dudal, P.~De Fabritiis, M.~S.~Guimaraes, I.~Roditi and S.~P.~Sorella,
%``Maximal violation of the Bell-Clauser-Horne-Shimony-Holt inequality via bumpified Haar wavelets,''
Phys. Rev. D \textbf{108} (2023), L081701
doi:10.1103/PhysRevD.108.L081701
[arXiv:2307.04611 [hep-th]].
%14 citations counted in INSPIRE as of 18 Mar 2026

%\cite{Dudal:2024bmf}
\bibitem{Dudal:2024bmf}
D.~Dudal and K.~Vandermeersch,
%``Further Evidence for Near-Tsirelson Bell-CHSH Violations in Quantum Field Theory via Haar Wavelets,''
[arXiv:2410.13362 [math-ph]].
%0 citations counted in INSPIRE as of 18 Mar 2026

%\cite{Dudal:2026eil}
\bibitem{Dudal:2026eil}
D.~Dudal and K.~Vandermeersch,
%``Near-Tsirelson Bell-CHSH Violations in Quantum Field Theory via Carleman and Hankel Operators,''
[arXiv:2604.05109 [math-ph]].
%2 citations counted in INSPIRE as of 25 May 2026


%\cite{Haag:1992hx}
\bibitem{Haag:1992hx}
R.~Haag, {\it Local Quantum Physics: Fields, Particles, Algebras}, Springer-Verlag (1992)
%110 citations counted in INSPIRE as of 06 Feb 2026





%\cite{Cirelson:1980ry}
\bibitem{Cirelson:1980ry}
B.~S.~Cirelson,
%``QUANTUM GENERALIZATIONS OF BELL'S INEQUALITY,''
Lett. Math. Phys. \textbf{4} (1980), 93-100
doi:10.1007/BF00417500
%1318 citations counted in INSPIRE as of 06 Feb 2026


\bibitem{Guimaraes:2025vfu}
M.~S.~Guimaraes, I.~Roditi and S.~P.~Sorella,
%``Class of bounded Hermitian operators for the Bell-Clauser-Horne-Shimony-Holt inequality in quantum field theory,''
Phys. Rev. D \textbf{112} (2025) no.8, 085009
doi:10.1103/x939-vzq7
[arXiv:2506.00504 [quant-ph]].
%2 citations counted in INSPIRE as of 06 Feb 2026

%\cite{Coleman:1985rnk}
\bibitem{Coleman:1985rnk}
S.~Coleman,
%``Aspects of Symmetry: Selected Erice Lectures,''
Cambridge University Press, 1985,
ISBN 978-0-521-31827-3
doi:10.1017/CBO9780511565045
%239 citations counted in INSPIRE as of 18 Mar 2026


\bibitem{Araki} 
H. Araki, J. Math. Phys. 5 (1964), 1-13

\bibitem{Rieffel} 
M. Rieffel and A. van Daele, Pac. J. Math. 69, 187 (1977)


%\cite{Guimaraes:2024alk}
\bibitem{Guimaraes:2024alk}
M.~S.~Guimaraes, I.~Roditi and S.~P.~Sorella,
%``Bell-CHSH inequality and unitary operators,''
Nucl. Phys. B \textbf{1008} (2024), 116717
doi:10.1016/j.nuclphysb.2024.116717
[arXiv:2403.15276 [quant-ph]].
%8 citations counted in INSPIRE as of 15 Feb 2026

%\cite{Banaszek:1998wcc}
\bibitem{Banaszek:1998wcc}
K.~Banaszek and K.~W{\'o}dkiewicz,
%``Testing Quantum Nonlocality in Phase Space,''
Phys. Rev. Lett. \textbf{82} (1999) no.10, 2009
doi:10.1103/PhysRevLett.82.2009
[arXiv:quant-ph/9806068 [quant-ph]].
%265 citations counted in INSPIRE as of 06 Mar 2026

%\cite{Guimaraes:2026xnu}
\bibitem{Guimaraes:2026xnu}
M.~S.~Guimaraes, I.~Roditi and S.~P.~Sorella,
%``Cat states and violation of the Bell-CHSH inequality in relativistic Quantum Field Theory,''
[arXiv:2601.05216 [hep-th]].
%0 citations counted in INSPIRE as of 09 Feb 2026




%\cite{Brunetti:1992zf,Borchers:1998ye}

%\cite{Brunetti:1992zf}
\bibitem{Brunetti:1992zf}
R.~Brunetti, D.~Guido and R.~Longo,
%``Modular structure and duality in conformal quantum field theory,''
Commun. Math. Phys. \textbf{156} (1993), 201-220
doi:10.1007/BF02096738
[arXiv:funct-an/9302008 [math.FA]].
%161 citations counted in INSPIRE as of 21 Mar 2026

%\cite{Borchers:1998ye}
\bibitem{Borchers:1998ye}
H.~J.~Borchers and J.~Yngvason,
%``Modular groups of quantum fields in thermal states,''
J. Math. Phys. \textbf{40} (1999), 601-624
doi:10.1063/1.532678
[arXiv:math-ph/9805013 [math-ph]].
%50 citations counted in INSPIRE as of 21 Mar 2026

%\cite{Foit:1983pbi,Mund:2005cv}

%\cite{Foit:1983pbi}
\bibitem{Foit:1983pbi}
J.~J.~Foit,
%``ABSTRACT TWISTED DUALITY FOR QUANTUM FREE FERMI FIELDS,''
Publ. Res. Inst. Math. Sci. Kyoto \textbf{19} (1983), 729-741
doi:10.2977/prims/1195182448
%15 citations counted in INSPIRE as of 22 Mar 2026

%\cite{Mund:2005cv}
\bibitem{Mund:2005cv}
J.~Mund, B.~Schroer and J.~Yngvason,
%``String-localized quantum fields and modular localization,''
Commun. Math. Phys. \textbf{268} (2006), 621-672
doi:10.1007/s00220-006-0067-4
[arXiv:math-ph/0511042 [math-ph]].
%86 citations counted in INSPIRE as of 22 Mar 2026

\end{thebibliography}
\end{document}